\documentclass[a4paper,12pt]{elsarticle}

\usepackage[numbers]{natbib}
\journal{ArXiv}
\usepackage{amsmath,amsfonts,amssymb}
\usepackage{amssymb}
\usepackage{mathrsfs}
\usepackage{lipsum}
\usepackage{float}
\usepackage{graphicx}
\usepackage{subcaption}
\usepackage{float} 
\newtheorem{remark}{Remark}[section]

\def\tsc#1{\csdef{#1}{\textsc{\lowercase{#1}}\xspace}}
\tsc{WGM}
\tsc{QE}
\tsc{EP}
\tsc{PMS}
\tsc{BEC}
\tsc{DE}

\newcommand{\UU}{\boldsymbol{U}}

\newcommand{\bu}{\boldsymbol{u}}
\newcommand{\bv}{\boldsymbol{v}}

\def\[{\left[}
\def\]{\right]}
\def\<{\langle}
\def\>{\rangle}
\def\({\left(}
\def\){\right)}
\def\[{\left [}
\def\]{\right]}
\def\({\left(}
\def\){\right)}

\begin{document}
	
\begin{frontmatter}
\renewcommand\arraystretch{1.0}
	
	\title{\textbf{A term-by-term variational multiscale method with dynamic subscales for incompressible turbulent aerodynamics}}
	
	\author{
		{\bf D.~Escobar} $^{1,2}$, \ 
		{\bf D.~Pacheco} $^{3,4,5}$, \
		{\bf A.~Aguirre} $^{1,2}$
		and \  
		{\bf E.~Castillo} $^{1,2,}$
		\footnote[0]{{\sf Email address:} {\tt ernesto.castillode@usach.cl}, 
			{\sf corresponding author}}\\
		{\small ${}^{1}$ Computational Heat and Fluid Flow Lab, Universidad de Santiago de Chile} \\
		{\small ${}^{2}$ Department of Mechanical Engineering, University of Santiago de Chile, Santiago, Chile} \\
		{\small ${}^{3}$ Chair for Computational Analysis of Technical Systems, RWTH Aachen University, Germany}\\
		{\small ${}^{4}$ Chair of Methods for Model-based Development in Computational Engineering, RWTH}\\
		{\small ${}^{5}$ Center for Simulation and Data Science (JARA-CSD), RWTH Aachen University, Germany}}

\begin{abstract}
Variational multiscale (VMS) methods offer a robust framework for handling under-resolved flow scales without resorting to problem-specific turbulence models. Here, we propose and assess a dynamic, term-by-term VMS stabilized formulation for simulating incompressible flows from laminar to turbulent regimes. The method is embedded in an incremental pressure-correction fractional-step framework and employs a minimal set of stabilization terms, yielding a unified discretization that (i) allows equal-order velocity--pressure interpolation and (ii) provides robust control of convection-dominated dynamics in complex three-dimensional settings. Orthogonal projections are a key ingredient and ensure that the non-residual, term-by-term structure induces dissipation through dynamic subscales suitable for turbulent simulations. The methodology is validated on large-scale external-aerodynamics configurations, including the Ahmed body at Re $ = 7.68\times 10^{5}$ for multiple slant angles, using unstructured tetrahedral meshes ranging from 3 to 40 million elements. Applicability is further demonstrated on a realistic Formula~1 configuration at $U_\infty=56~\mathrm{m/s}$ (201.6~km/h), corresponding to Re $ \approx 10^{6}$. The results show that the proposed stabilized pressure-segregated formulation remains robust at scale and captures key separated-flow features and coherent wake organization. Pointwise velocity and pressure spectra provide an a posteriori consistency indicator, exhibiting finite frequency ranges compatible with inertial-subrange reference slopes in the resolved band and supporting dissipation control in under-resolved regimes within a unified stabilized finite element framework.
\end{abstract}

\end{frontmatter}


\section{Introduction}
Accurate turbulence representation is central to computational aerodynamics as it controls drag, lift and separation, while also governing unsteady loads relevant to structural integrity under adverse operating conditions. In realistic configurations, the broad range of spatial and temporal scales renders direct numerical simulation (DNS) computationally prohibitive at engineering Reynolds numbers \cite{li2022assessment,mani2023perspective}. Conversely, Reynolds-averaged Navier--Stokes (RANS) closures may exhibit limited predictive capability and strong case dependence in the presence of complex geometry and separated flow \cite{AULTMAN2022105297,SHUKLA2021108954,GHIDONI2024105881}. This motivates turbulence-resolving numerical strategies that remain tractable on practical meshes while capturing boundary-layer evolution, wake dynamics, and vortex interactions.

Large-eddy simulation (LES) provides a widely used intermediate modeling level by resolving the large, energy-containing structures and modeling the smallest scales \cite{bouhafid2024combined}. In external aerodynamics, LES can improve predictions of unsteady separated flows and coherent wake dynamics compared with typical RANS practice, but its cost remains substantial for high-Reynolds conditions and realistic geometries, often restricting its use to simplified configurations, well-established benchmarks, or localized high-fidelity studies within hybrid workflows \cite{janocha2022large,kim2024large,shen2025comparative,teng2025atmospheric}. These constraints have motivated continued research on discretizations and modeling strategies that control numerical dissipation while preserving robustness in under-resolved regimes, including high-order incompressible LES approaches for fully inhomogeneous turbulence \cite{SHETTY20108802} and low-dissipation formulations that leverage fractional-step strategies to enable stable computations with practical mixed or equal-order velocity--pressure spaces \cite{LEHMKUHL201951}. In parallel, hybrid RANS/LES strategies and alternative discretization frameworks continue to be explored to reduce cost while retaining fidelity in complex wall-bounded flows and vehicle-like configurations \cite{MOZAFFARI2024113269}. More recently, attention has also focused on heterogeneous turbulent environments and geometrical flexibility, where the combined choice of convection schemes and subgrid-scale closures becomes critical for dissipation control and accuracy \cite{AJAY2026114554}.

In this context, \emph{implicit LES} emerged as a modeling framework where the numerical method itself approximates turbulent dissipation, instead of requiring explicit, physical turbulence modeling assumptions. Among such implicit techniques, MILES \cite{Margolin2002,Margolin2006} and VMS \cite{Ahmed2017} methods are the prominent examples in the finite-volume and finite-element communities, respectively. For an in-depth discussion on the important connections between physical and numerical dissipation in turbulent flows, we refer the reader to the recent work of \citet{Fehn2025}. Within this landscape, the VMS framework offers a systematic approach for incompressible flow simulation and turbulence modeling, based on a decomposition into resolvable scales and unresolved subscales whose effect is retained at the discrete level \cite{COLOMES2015,Ahmed2017,Rasthofer2018}. In turbulent settings, dynamic subscales provide a mechanism to embed controlled numerical dissipation and recover implicit LES behavior without introducing explicit eddy-viscosity closures \cite{COLOMES2015,GRAVEMEIER2010853}. Comparative studies across well-established turbulent benchmarks have shown that modeling and implementation choices, including the definition of the subscale space, the use of static versus dynamic subscales, and the enforcement of orthogonality, directly influence stability, dissipation properties, and computational efficiency \cite{COLOMES2015,OSSES2021114099,AHMED2020112997}. More mathematically oriented contributions have also investigated robust formulations and convergence properties for incompressible flow discretizations augmented with VMS-LES stabilization, including treatments suitable for complex boundaries and slip conditions \cite{bansal2024nitsche}. Beyond single-physics turbulence, VMS methods have also been explored in coupled contexts such as fluid--structure interaction and aeroelasticity \cite{SVACEK2023115125}, and recent work combines VMS-based turbulence modeling with adaptive and space--time strategies to extend applicability to high-Reynolds-number flows on anisotropic and evolving meshes \cite{ravensbergen2020variational,TEMELLINI2025114123}.

A central development within the VMS framework are stabilized finite element formulations based on orthogonal and dynamic subscales \cite{CODINA20001579,Codina2007,Badia2009}. Enforcing orthogonality between resolved and unresolved spaces provides a consistent scale separation, enables stable equal-order velocity--pressure interpolation, and yields a physically meaningful definition of numerical dissipation \cite{PRINCIPE2010791,GUASCH2013154}. When combined with time-dependent subscales, this approach removes restrictive time-step constraints, improves robustness under anisotropic space--time discretizations, and recovers key features of turbulent energy transfer, including inertial-range dissipation and backscatter effects \cite{Codina2007,PRINCIPE2010791}. Beyond residual-based realizations, term-by-term stabilized formulations based on orthogonal projections preserve optimal convergence properties while providing a clear dissipative structure in which the contribution of individual stabilization terms can be identified \cite{CASTILLO2019701,GONZALEZ20201009}. This structural clarity has enabled systematic energy-budget analyses and has clarified how numerical dissipation arises from specific subscale components rather than case-dependent closure assumptions \cite{GUASCH2013154,CHEN2022114280}. Importantly, these concepts are not restricted to Newtonian fluids, and orthogonal dynamic subscale formulations have been extended to more general rheological settings, supporting the flexibility of the VMS paradigm across a broad class of flow problems \cite{AMANI2021104615,GONZALEZ2022115586,KATZ2023115036}.

Despite these advances, much of the numerical evidence available for VMS-based turbulence modeling still relies on relatively simple benchmark geometries, such as homogeneous isotropic turbulence, Taylor--Green vortices, and turbulent channel flows \cite{COLOMES2015,AHMED2020112997,TEMELLINI2025114123}. While these configurations are essential to analyze dissipation mechanisms and turbulence statistics under controlled conditions, large-scale validation in complex external aerodynamics settings remains comparatively less documented, even though separation, wake interactions, and multi-scale vortical structures impose stringent requirements on stabilization design, dissipation control, and numerical robustness. Moreover, industrially relevant computations often rely on fractional-step (i.e., pressure-segregation) strategies, including pressure-correction schemes, to enable efficient solution of the incompressible Navier--Stokes equations on large unstructured meshes. There is therefore limited large-scale evidence on how VMS-based turbulence-resolving formulations behave within pressure-correction fractional-step schemes for complex, under-resolved external-aerodynamics flows.

To bridge the gap between controlled benchmarks and application-driven external aerodynamics, this work considers two complementary configurations. The Ahmed body is a well-established benchmark for assessing turbulence-resolving methodologies for incompressible external aerodynamics, as it retains the essential flow physics of road vehicles while keeping the geometry sufficiently controlled for systematic comparisons \cite{ahmed1984wake,aljure2014turbulent}. A distinctive feature is the strong sensitivity of the wake topology and aerodynamic drag to the rear slant angle, which induces markedly different separation and vortex-interaction regimes \cite{SERRE201310,liu2021flow}. This sensitivity has motivated extensive experimental and numerical investigations across slant angles and has fostered collaborative evaluation efforts that provide reference databases for eddy-resolving simulations, including LES and DES at $\mathrm{Re} = 7.68 \times 10^{5}$ \cite{SERRE201310}. Beyond time-averaged quantities, the Ahmed-body wake exhibits rich unsteady dynamics, including bistable and turbulence-intensity-dependent behavior in square-back configurations \cite{burton2021influence,sagharichi2025effects}. Therefore, the Ahmed body continues to serve as a key validation platform for modern high-fidelity and hybrid approaches, as well as for modeling ingredients such as wall treatments and immersed-boundary strategies under massive separation \cite{cai2022application,GHIDONI2024105881,AGUERRE2024105635}.

As a second and more demanding case, we consider a realistic Formula~1 car configuration constrained by FIA regulations, where aerodynamic performance depends on the interaction of multiple lifting surfaces and strongly three-dimensional flow structures \cite{FIA2025TechRegs}. In contrast to simplified road-vehicle geometries, the flow around an F1 car features a substantially more intricate topology, including front and rear wings, multi-element aerodynamic devices, and complex wheel and bodywork interactions, together with a higher operational Reynolds numbers and a broader range of separation- and vortex-dominated mechanisms. Recent benchmark-oriented studies on regulation-inspired F1 front-wing geometries operating under ground effect, supported by publicly available CAD models and experimental datasets, have highlighted the resolution requirements and validation challenges associated with capturing the dominant flow topology and loading trends \cite{BUSCARIOLO2022104832}. Complementary comparisons between explicit and implicit LES strategies on related front-wing-in-ground-effect settings further emphasize the role of discrete stability and dissipation control in under-resolved high-fidelity simulations \cite{NTOUKAS2025104425}. This configuration therefore provides a demanding large-scale setting to assess robustness, dissipation control, and scalability of turbulence-resolving discretizations on complex unstructured meshes.

In this work, we propose and assess a dynamic, term-by-term VMS-inspired stabilized formulation tailored to an incremental pressure-correction fractional-step method for incompressible flows. The formulation is constructed with a minimal stabilization layout, in the sense that it introduces only the terms required to enable stable equal-order velocity--pressure interpolation and control convection in turbulent regimes, while preserving a clear dissipation mechanism through orthogonal dynamic subscales. A key design aspect is that stabilization terms are embedded directly into the fractional-step stages by modifying the corresponding operator blocks, without introducing additional velocity--pressure coupling. In a projection setting, such cross terms would typically enter as right-hand-side contributions, which do not enhance stability and may effectively reintroduce coupling across stages. Time integration is handled consistently by applying BDF2 both to the resolved variables and to the dynamic subscales, yielding a fully second-order temporal framework. An optional grad--div contribution acting through the pressure subscale is also considered; while not required by the baseline formulation, it improves nonlinear robustness in demanding configurations \cite{Rhoe2010}.

Our methodology is validated on challenging aerodynamic configurations that combine turbulence, separation, and complex geometries, including the Ahmed body for multiple slant angles using unstructured tetrahedral meshes with up to 40 million elements. Applicability is further demonstrated on the F1 car at a representative on-track mean velocity of $200~\mathrm{km/h}$, corresponding to $\mathrm{Re} \approx 1.13 \times 10^{6}$. In both cases, integral quantities such as drag coefficients are complemented with pointwise velocity and pressure spectra as practical diagnostics of dissipation control and robustness in large-scale under-resolved simulations.

The remainder of this article is organized as follows. Section~\ref{sec:gov_eq} introduces the governing equations and the variational setting. Section~\ref{sec:fractional_step} summarizes the incremental pressure-correction time integration. Section~\ref{sec:stab} presents the proposed orthogonal, term-by-term stabilized formulation and its block-wise embedding into the fractional-step stages. Section~\ref{sec:ahmed_problem} reports the Ahmed-body setup and results, and Section~\ref{sec:f1} presents the Formula~1 application. Finally, Section~\ref{Conclusion} summarizes the main findings and outlines directions for future work.

\section{Governing equations}
\label{sec:gov_eq}

Let $\Omega\subset\mathbb{R}^d$ ($d=2,3$) be a bounded computational domain with boundary $\Gamma$ and final time $T>0$.
We consider incompressible flow of a Newtonian fluid with constant density $\rho$ and dynamic viscosity $\mu$.
The unknowns are the velocity $\bu(\boldsymbol{x},t)$ and pressure $p(\boldsymbol{x},t)$, governed by
\begin{align}
\rho\partial_t \bu + (\rho\bu\cdot\nabla)\bu - \mu\Delta \bu + \nabla p &= \boldsymbol{f}
\quad \text{in }\Omega\times(0,T], \label{eq:NS_mom}\\
\nabla\cdot \bu &= 0
\quad \text{in }\Omega\times(0,T]. \label{eq:NS_inc}
\end{align}

The boundary is decomposed as $\Gamma=\Gamma_D\cup\Gamma_N$, $\Gamma_D\cap\Gamma_N=\emptyset$.
We prescribe Dirichlet, outflow and initial conditions
\begin{align}
\bu &= \bu_g
\quad \text{on }\Gamma_D\times(0,T], \label{eq:BC_D}\\
(\mu\nabla \bu - p\boldsymbol{I})\boldsymbol{n} &=  \mathbf{0}
\quad \text{on }\Gamma_N\times(0,T], \label{eq:BC_N}\\
\bu  &=  \bu_0
\quad \text{at }t=0, 
\end{align}
where $\boldsymbol{I}$ denotes the $d\times d$ identity tensor and $\boldsymbol{n}$ is the unit outward normal. Condition \eqref{eq:BC_N} is written in pseudo-traction form, consistent with the outflow condition used in the benchmark setups.
To simplify the exposition in the next sections, we restrict the presentation to homogeneous Dirichlet conditions in what follows, i.e.,
$\bu_g=\boldsymbol{0}$ on $\Gamma_D$.

The aerodynamic configurations considered in this work are reported under wind-tunnel-like conditions commonly adopted in the experimental
literature. Accordingly, $U_\infty$ denotes the prescribed free-stream inflow speed, and reference areas and lengths are selected according
to the standard definitions of each benchmark.

\subsection{Aerodynamic forces and coefficients}
\label{subsec:forces}

Let $\Gamma_i\subset\Gamma_D$ denote the body surface where forces are evaluated.
The total force is
\begin{equation}
\boldsymbol{F} = \int_{\Gamma_i} \boldsymbol{\sigma}\,\boldsymbol{n} \,\mathrm{d}\Gamma_i,
\qquad
\boldsymbol{\sigma} := -p\boldsymbol{I} + \mu\nabla\bu + (\mu\nabla\bu)^{\top},
\label{eq:force_def}
\end{equation}
where $\boldsymbol{\sigma}$ is the Cauchy stress tensor.
The drag force $F_D$ is the component of $\boldsymbol{F}$ in the direction of $U_\infty$, while the lift force $F_L$ is the component
orthogonal to it. The corresponding dimensionless coefficients are
\begin{equation}
C_D = \frac{F_D}{\tfrac{1}{2}\rho U_\infty^2 A},
\qquad
C_L = \frac{F_L}{\tfrac{1}{2}\rho U_\infty^2 A},
\label{eq:CD_CL}
\end{equation}
where $A$ is a benchmark-specific reference area (e.g., frontal area). The Reynolds number is defined as
\begin{equation}
\mathrm{Re} = \frac{\rho l U_\infty}{\mu},
\label{eq:Re}
\end{equation}
with $l$ being a characteristic length specified for each flow problem.


\subsection{Variational setting, discretization, and block notation}
\label{subsec:variational_discretization}

We summarize the variational setting and the fully discrete Galerkin framework employed throughout the paper, also introducing a block-matrix
notation that will be used as a means of locating the stabilized contributions \emph{term-by-term} in the monolithic operator.

Let $V:=[H^1_0(\Omega)]^d$ and $Q:=L^2(\Omega)$, endowed with the $L^2(\Omega)$ inner product $\langle \cdot,\cdot\rangle$.
A standard weak formulation of \eqref{eq:NS_mom}--\eqref{eq:NS_inc} is: find $(\bu,p):(0,T]\to V\times Q$ such that
\begin{equation}
\left\langle \rho\,\partial_t \bu, \bv \right\rangle
+ B(\bu; \bu,p; \bv,q)
= \left\langle \boldsymbol{f}, \bv \right\rangle
\qquad \forall (\bv,q)\in V\times Q,
\label{eq:weak}
\end{equation}
where
\begin{equation}
B(\hat{\bu}; \bu,p; \bv,q)
:= \left\langle \rho\hat{\bu}\cdot\nabla \bu, \bv \right\rangle
+ \left\langle \mu\nabla \bu, \nabla \bv \right\rangle
- \left\langle p, \nabla\cdot \bv \right\rangle
+ \left\langle \nabla\cdot \bu, q \right\rangle ,
\label{eq:form_B}
\end{equation}
and $\hat{\bu}$ denotes a convective velocity used for linearization (e.g., a Picard strategy).

Let $\mathcal{T}_h$ be a conforming triangulation of $\Omega$ and $V_h\subset V$ and $Q_h\subset Q$ be conforming finite element spaces. The Galerkin approximation reads: find $(\bu_h,p_h):(0,T]\to V_h\times Q_h$ such that
\begin{equation}
\left\langle \rho\partial_t \bu_h, \bv_h \right\rangle
+ B(\bu_h; \bu_h,p_h; \bv_h,q_h)
= \left\langle \boldsymbol{f}, \bv_h \right\rangle
\qquad \forall (\bv_h,q_h)\in V_h\times Q_h.
\label{eq:galerkin}
\end{equation}
In the turbulent computations considered here, Eqs.~\eqref{eq:weak}--\eqref{eq:galerkin} provide the mathematical template for the
method, while the discrete solution is sought in under-resolved regimes where robustness and controlled numerical dissipation are essential
to capture the dominant flow structures and the relevant integral observables (e.g., forces and spectra).

Time integration is performed with the second-order backward differentiation formula (BDF2): for a generic quantity $\boldsymbol{v}_h(t)$,
\begin{equation}
\partial_t \boldsymbol{v}_h(t^{n+1})
= \alpha_0 \boldsymbol{v}_h^{n+1} - \alpha_1 \boldsymbol{v}_h^{n} + \alpha_2 \boldsymbol{v}_h^{n-1}
+ \mathcal{O}(\Delta t^2),
\qquad
\alpha_0=\frac{3}{2\Delta t},\;
\alpha_1=\frac{2}{\Delta t},\;
\alpha_2=\frac{1}{2\Delta t}.
\label{eq:bdf2}
\end{equation}

\paragraph{Fully discrete block form (notation).}
To expose the saddle-point structure that will later be modified \emph{term-by-term} by the stabilization, we introduce a fully discrete
block representation of the Galerkin system at time $t^{n+1}$. This notation is used only to identify which operator blocks are affected
by each stabilized contribution.

Let $\boldsymbol{M}$ denote the velocity mass matrix, $\boldsymbol{K}$ the discrete diffusion operator,
$\boldsymbol{C}(\hat{\bu}_h)$ the discrete convection operator, and $\boldsymbol{B}$ the discrete divergence operator, with
$\boldsymbol{B}^\top$ the corresponding discrete gradient. Then, the coupled Galerkin system can be written formally as
\begin{equation}
\underbrace{\begin{bmatrix}
\alpha_0 \boldsymbol{M} & \boldsymbol{0}\\
\boldsymbol{0} & \boldsymbol{0}
\end{bmatrix}}_{\text{time (mass) block}}
\begin{bmatrix}
\bu^{n+1}\\ p^{n+1}
\end{bmatrix}
+
\underbrace{\begin{bmatrix}
\boldsymbol{C}(\hat{\bu}_h) + \boldsymbol{K} & \boldsymbol{B}^\top\\
\boldsymbol{B} & \boldsymbol{0}
\end{bmatrix}}_{\displaystyle \mathcal{K}}
\begin{bmatrix}
\bu^{n+1}\\ p^{n+1}
\end{bmatrix}
=
\begin{bmatrix}
\boldsymbol{b}^{\,n+1}\\ \boldsymbol{0}
\end{bmatrix},
\label{eq:galerkin_block}
\end{equation}
where $\boldsymbol{b}^{\,n+1}$ collects known BDF2 history terms, boundary-condition contributions, and external forces.
This block structure will be used in Section~\ref{sec:stab} to pinpoint where each stabilization term enters and how it is embedded into the
incremental pressure-correction framework while preserving the computational organization of the projection method.

\section{Fractional-step time integration}
\label{sec:fractional_step}
The numerical simulation of incompressible turbulent flows typically requires advancing the solution over a very large number of time steps
to capture the relevant unsteady dynamics. This challenge is intensified by the three-dimensional nature of turbulence and the large number
of degrees of freedom associated with realistic aerodynamic configurations. In this context, fractional-step (projection) methods offer an
efficient alternative to fully coupled velocity--pressure formulations, as they lead to linear subproblems with substantially reduced computational complexity. In this work, we employ an \emph{incremental pressure-correction} scheme combined with BDF2 time integration. Given $(\bu^n,p^n)$ and $(\bu^{n-1},p^{n-1})$, the algorithm advances the solution to $t^{n+1}=t^n+\Delta t$ as follows.

\paragraph{Step 1: Velocity predictor.}
Compute an intermediate (non-solenoidal) velocity $\bu^*$ by solving, for all $\bv_h\in V_h$,
\begin{equation}
\left\langle
\rho\frac{3\bu^* - 4\bu^n + \bu^{n-1}}{2\Delta t},
\bv_h
\right\rangle
+
\left\langle
\rho\hat{\bu}_h \cdot \nabla \bu^*,
\bv_h
\right\rangle
+
\left\langle
\mu \nabla \bu^*,
\nabla \bv_h
\right\rangle
-
\left\langle
p^n,\nabla \cdot \bv_h
\right\rangle
=
\left\langle
\boldsymbol{f}^{n+1},\bv_h
\right\rangle.
\label{eq:velocity_predictor}
\end{equation}
The convective velocity is treated by Picard linearization. Specifically, within the Picard loop we set
$\hat{\bu}_h=\bu^{*,(i-1)}$, where $(i-1)$ denotes the previous nonlinear iterate, yielding a robust fixed-point treatment of the convective term.

\paragraph{Step 2: Pressure correction.}
Compute the pressure increment $\delta p^{\,n+1} := p^{n+1}-p^n$ by solving, for all $q_h\in Q_h$,
\begin{equation}
\left\langle
\nabla \delta p^{\,n+1},
\nabla q_h
\right\rangle
=
\frac{3\rho}{2\Delta t}
\left\langle
\nabla \cdot \bu^*,
q_h
\right\rangle.
\label{eq:pressure_poisson}
\end{equation}
Pressure boundary conditions are set as $\delta p^{\,n+1}=0$ on $\Gamma_N$ and $\boldsymbol{n}\cdot\nabla \delta p^{\,n+1}=0$ on $\Gamma_D$ \cite{Guermond2006}.

\paragraph{Step 3: Velocity correction and pressure update.}
Obtain the (weakly) divergence-free velocity through
\begin{equation}
\left\langle
\bu^{n+1},
\bv_h
\right\rangle
=
\left\langle
\bu^*,
\bv_h
\right\rangle
-
\frac{2\Delta t}{3\rho}
\left\langle
\nabla \delta p^{\,n+1},
\bv_h
\right\rangle,
\label{eq:velocity_correction}
\end{equation}
and update the pressure as
\begin{equation}
p^{n+1} = p^n + \delta p^{\,n+1}.
\label{eq:pressure_update_fs}
\end{equation}

This incremental formulation can yield up to second-order accuracy in time when combined with BDF2; an overview of classical projection methods and related schemes can be found in the review by \citet{Guermond2006}.
\section{Stabilized formulation with finite elements}
\label{sec:stab}

We shall introduce a stabilized finite element formulation designed to operate in turbulent regimes on unstructured meshes, while remaining compatible with a pressure-correction fractional-step strategy. The stabilization is formulated at the fully discrete monolithic level and is embedded into the fractional-step stages through a term-by-term construction, so that the computational
structure of the projection method is preserved.

\subsection{Orthogonal VMS viewpoint and motivation}
\label{subsec:vms_motivation}

Equal-order finite element spaces are adopted for velocity and pressure due to their simplicity and flexibility on large-scale unstructured
discretizations. Such pairs, however, require stabilization to control the pressure--velocity coupling and to maintain robustness in
convection-dominated regimes. To this end, we employ an \emph{orthogonal variational multiscale} (VMS) approach, in which unresolved
information is sought in the $L^2$-orthogonal complement of the finite element space.

The VMS framework is based on a scale decomposition of the solution into a resolved component and an unresolved correction (subscale). For a
generic unknown $\UU$ (or a collection of unknowns), we decompose
\begin{equation}
\UU = \UU_h + \widetilde{\UU},
\ \, \text{where}
\ \
\UU_h\in \boldsymbol{\mathcal{X}}_h \; \text{and}
\ \
\widetilde{\UU}\in \widetilde{\boldsymbol{\mathcal{X}}} 
,
\label{eq:vms_split}
\end{equation}
with $\boldsymbol{\mathcal{X}}_h$ and $\widetilde{\boldsymbol{\mathcal{X}}}$ denoting the finite element space and its complement, respectively ($\boldsymbol{\mathcal{X}} = \boldsymbol{\mathcal{X}}_h \oplus \widetilde{\boldsymbol{\mathcal{X}}}$). The orthogonal
projector $P_h^\perp:\boldsymbol{\mathcal{X}}\to\widetilde{\boldsymbol{\mathcal{X}}}$ is defined by
\begin{equation}
\left\langle P_h^\perp \phi, v_h \right\rangle = 0
\qquad \forall v_h \in \boldsymbol{\mathcal{X}}_h .
\label{eq:orth_proj}
\end{equation}
This operator enforces a strict scale separation: the finite element space captures the resolved field, while the subscales carry only the
unresolved corrections. Such an orthogonal decomposition avoids redundancy between scales and yields a formulation in which each
stabilization contribution can be directly associated with a specific operator acting on the resolved variables.

\subsection{Term-by-term stabilized formulation at the monolithic level}
\label{subsec:monolithic_stab}

The stabilized formulation is first posed at the coupled, fully discrete level and is subsequently embedded into the incremental pressure-correction scheme. The design principle is to preserve the structure of the projection method: stabilization terms are introduced \emph{term-by-term} so that no additional velocity--pressure cross couplings are created. Crucially, we avoid cross-stabilization terms that mix pressure and velocity in a manner incompatible with pressure segregation. In a pressure-correction method, such couplings typically appear as additional right-hand-side contributions in the decoupled
subproblems; they do not act as stabilizing operators within the corresponding stage and may effectively reintroduce coupling across fractional-step stages.

The stabilized problem consists in finding
$\boldsymbol{U}_h^{n+1}=[\bu_h^{n+1},p_h^{n+1}]\in V_h\times Q_h$ such that, for all
$\boldsymbol{V}_h=[\bv_h,q_h]\in V_h\times Q_h$,
\begin{equation}
\left\langle \rho\partial_t \bu_h, \bv_h \right\rangle
+ B(\hat{\bu}_h; \boldsymbol{U}_h,\boldsymbol{V}_h)
-\sum_K \big(\widetilde{p}, \nabla\cdot\bv_h\big)_K
-\sum_K \big(\widetilde{\bu}_1, \rho\,\hat{\bu}_h\cdot\nabla\bv_h\big)_K
-\sum_K \big(\nabla q_h, \widetilde{\bu}_2\big)_K
=
\left\langle \boldsymbol{f}, \bv_h \right\rangle ,
\label{eq:stab_weak}
\end{equation}
where $(\cdot,\cdot)_K$ denotes the elementwise $L^2$ inner product.

The pressure subscale is taken quasi-static,
\begin{equation}
\tau_2^{-1}(\hat{\bu}_h)\,\widetilde{p}
= -\,P_h^\perp(\nabla\cdot \bu_h),
\label{eq:sub_p}
\end{equation}
while the velocity subscales are dynamic and split into convective and pressure-gradient contributions,
\begin{align}
\rho\partial_t \widetilde{\bu}_1 + \tau_1^{-1}(\hat{\bu}_h)\,\widetilde{\bu}_1
&= -\,P_h^\perp\!\big(\rho\hat{\bu}_h\cdot\nabla \bu_h\big),
\label{eq:sub_u1}\\
\rho\partial_t \widetilde{\bu}_2 + \tau_1^{-1}(\hat{\bu}_h)\,\widetilde{\bu}_2
&= -\,P_h^\perp(\nabla p_h).
\label{eq:sub_u2}
\end{align}

Time integration of the velocity subscales is performed consistently (BDF2). At the fully discrete level,
\begin{align}
\left[\tau_1^{-1}(\hat{\bu}_h) + \frac{3\rho}{2\Delta t}\right]\widetilde{\bu}_1^{\,n+1}
&= \rho\frac{4\widetilde{\bu}_1^{\,n}-\widetilde{\bu}_1^{\,n-1}}{2\Delta t}
- P_h^\perp\!\big(\rho\hat{\bu}_h\cdot\nabla \bu_h^{n+1}\big),
\label{eq:sub_u1_bdf2}\\
\left[\tau_1^{-1}(\hat{\bu}_h) + \frac{3\rho}{2\Delta t}\right]\widetilde{\bu}_2^{\,n+1}
&= \rho\frac{4\widetilde{\bu}_2^{\,n}-\widetilde{\bu}_2^{\,n-1}}{2\Delta t}
- P_h^\perp(\nabla p_h^{n+1}).
\label{eq:sub_u2_bdf2}
\end{align}

The stabilization parameters are defined as
\begin{equation}
\tau_1(\hat{\bu}_h)
=
\left(
c_1\frac{\mu}{h_1^2}
+
c_2\frac{\rho|\hat{\bu}_h|}{h_2}
\right)^{-1},
\qquad
\tau_2(\hat{\bu}_h)
=
\frac{h_1^2}{c_3}\,\tau_1(\hat{\bu}_h),
\label{eq:tau12}
\end{equation}
where $h_1$ and $h_2$ are characteristic element lengths and $c_1,c_2,c_3$ are algorithmic constants. In this work we employ $c_1=4$,
$c_2=2$, and $c_3=24$. The choice $c_3\neq c_1$ departs from common settings in which the same constant is used for both parameters. Here,
$c_3=24$ was selected based on systematic numerical experimentation after establishing a convergent baseline for the F1 configuration,
and it provided the most robust nonlinear behavior across all test cases considered.

Through \eqref{eq:sub_p}, we obtain a grad--div-type contribution. In the F1 application, its inclusion led to a marked improvement in nonlinear robustness and enabled convergence on the available meshes and time-step sizes; without it, the
nonlinear iterations tended to stagnate. This behavior is consistent with an insufficient level of numerical dissipation in that
configuration, and the additional divergence-related stabilization provides an effective remedy in the most demanding case examined. When
higher-order elements are employed, numerical evidence further suggests that incorporating an explicit dependence on the polynomial degree
in the stabilization parameters may be beneficial \cite{Guerrero2025}.

\begin{remark}
The grad--div stabilisation has the form $\sum_{K}(\tau_2\nabla\cdot\boldsymbol{u}_h,\nabla\cdot\boldsymbol{v}_h)$, which is why it is sometimes also referred to as ``div--div''. Its more usual name comes from the fact that this term can be reinterpreted as originating, at the continuous (strong) level, from a consistently added grad(div) term $-\nabla(\tau_2\nabla\cdot\boldsymbol{u})$ \cite{John2016,He2025}.
\end{remark}
\subsection{Block structure and connection to the fractional-step stages}
\label{subsec:block_and_fs}

Using the fully discrete block notation introduced in Section~\ref{subsec:variational_discretization}, the stabilized monolithic system at time
level $t^{n+1}$ can be written as
\begin{equation}
\begin{bmatrix}
\alpha_0 \boldsymbol{M} & \boldsymbol{0}\\
\boldsymbol{0} & \boldsymbol{0}
\end{bmatrix}
\begin{bmatrix}
\bu^{n+1}\\ p^{n+1}
\end{bmatrix}
+
\underbrace{\begin{bmatrix}
\boldsymbol{C}(\hat{\bu}_h)+\boldsymbol{K}+\boldsymbol{H}_s & \boldsymbol{B}^\top\\
\boldsymbol{B} & \boldsymbol{L}_s
\end{bmatrix}}_{\displaystyle \mathcal{K}_s}
\begin{bmatrix}
\bu^{n+1}\\ p^{n+1}
\end{bmatrix}
=
\begin{bmatrix}
\boldsymbol{b}^{\,n+1}+\boldsymbol{b}_s^{\,n+1}\\
\boldsymbol{g}_s^{\,n+1}
\end{bmatrix}.
\label{eq:stab_block_separated}
\end{equation}
Here, $\boldsymbol{H}_s$ collects stabilization contributions acting in the momentum equation, while $\boldsymbol{L}_s$ denotes the
pressure-related diagonal block induced by the pressure-gradient subscale. The vectors $\boldsymbol{b}_s^{\,n+1}$ and
$\boldsymbol{g}_s^{\,n+1}$ collect known history terms originating from the BDF2 discretization of the dynamic subscales.

This block-level representation clarifies the interaction between stabilization and projection. When embedded into the incremental
pressure-correction framework, momentum-side stabilization enters the velocity predictor stage, whereas the pressure-related diagonal
contribution augments the pressure-correction Poisson solve. The remaining subscale-history terms appear only as known right-hand-side
contributions, so the final velocity projection step retains its standard form. This is consistent with classical stability analysis of projection-type finite element schemes \cite{CODINA2001112}, which shows that pressure control may deteriorate for small $\Delta t$ unless additional stabilization mechanisms are introduced, particularly for second-order variants. Within that perspective, the present construction provides pressure stability through $\boldsymbol{L}_s$ and
introduces the minimum additional momentum dissipation required for robustness through $\boldsymbol{H}_s$, while preserving the
standard projection structure.

\subsection{Implementation details}
\label{subsec:implementation}

The formulation involves nonlinearities arising from convection and from stabilization parameters depending on $\hat{\bu}_h$.
The main implementation aspects are as follows.
\begin{itemize}

\item \textbf{Linearization of convection.}
Convection is treated by a fixed-point (Picard) iteration. At each nonlinear iteration, we fix the convective velocity $\hat{\bu}_h$ (using its value at the previous iteration) and solve the resulting linear subproblems.

\item \textbf{Iterative realization of orthogonal projections.}
The orthogonal projector is realized via a projection--correction strategy. For a generic quantity $f$,
\begin{equation}
P_h^\perp(f) \approx f - P_h(f),
\end{equation}
where $P_h$ denotes the $L^2$ projector onto the discrete space. In practice, $P_h$ is updated within the Picard loop using the previous
iterate on the right-hand side, i.e.,
\begin{equation}
P_h^\perp\!\big(f^{(i)}\big)
=
f^{(i)} - P_h\!\big(f^{(i-1)}\big),
\end{equation}
where the projection $z_h:= P_h\!\big(f^{(i-1)}\big) \in \boldsymbol{X}_h\subset L^2(\Omega)$ is simply the solution of
\begin{align}
    \left\langle z_h, w_h \right\rangle = \left\langle f^{(i-1)}, w_h \right\rangle\label{massSolve}
\end{align}
for all $w_h\in \boldsymbol{X}_h$. This yields a modest overhead relative to the overall nonlinear iteration in large-scale computations. In this work we restrict attention to $L^2(\Omega)$ projections, but alternative strategies can be found in \cite{codina2018hp,GRAVENKAMP2024112754}.

\item \textbf{Mass-matrix solve.}
Each update of $P_h$ requires the solution of a mass-matrix problem \eqref{massSolve}. For first-order elements, a lumped mass matrix provides an efficient approximation, while for higher-order discretizations a consistent mass matrix is usually required to preserve optimal accuracy.
\end{itemize}


\section{Problem definition: Ahmed body}
\label{sec:ahmed_problem}

We first validate the proposed dynamic, term-by-term VMS-inspired stabilization on the Ahmed body, a reference configuration in external
aerodynamics. This benchmark concentrates flow mechanisms that stress turbulence-resolving discretizations on practical meshes: boundary-layer
separation over the rear slant, large recirculation regions, and coherent vortical structures in the wake. Moreover, wake topology and
aerodynamic drag depend strongly on the rear slant angle, making the Ahmed body a stringent test for stabilized formulations in separated
three-dimensional flows.

\subsection{Geometry, flow conditions, and computational domain}
\label{subsec:ahmed_setup}

The Ahmed-body geometry consists of a rectangular main body with a rear slanted surface with angle $\theta$ (Fig.~\ref{fig:ahmed_geometry}),where the main geometric features and characteristic dimensions are indicated  \cite{ahmed1984wake}, which strongly influences separation and wake organization. We consider
$\theta \in \{0^\circ,12.5^\circ,20^\circ,25^\circ,35^\circ\}$ at $\mathrm{Re} = 7.68\cdot 10^{5}$, based on the body length $L = 1.044~\mathrm{m}$ and the inlet velocity $U_\infty = 40~\mathrm{m/s}$.
All simulations are carried out under wind-tunnel-like conditions consistent with standard experimental setups commonly reported for this
benchmark.

The computational domain extends $3L$ upstream of the body, $5L$ downstream, and $2L$ in the lateral and vertical directions (Fig.~\ref{fig:ahmed_domain}),where the domain extents relative to the body length are reported, limiting the
influence of far-field boundaries on the separated wake. A uniform inflow velocity is prescribed at the inlet, and no-slip conditions are
enforced on the body surface and on the ground. Far-field boundaries are treated with traction/pressure-type external-flow conditions,
consistent with the reference literature and with the pseudo-stress traction form adopted in Section~\ref{sec:gov_eq}.

\begin{center}
    \includegraphics[width=0.7\linewidth]{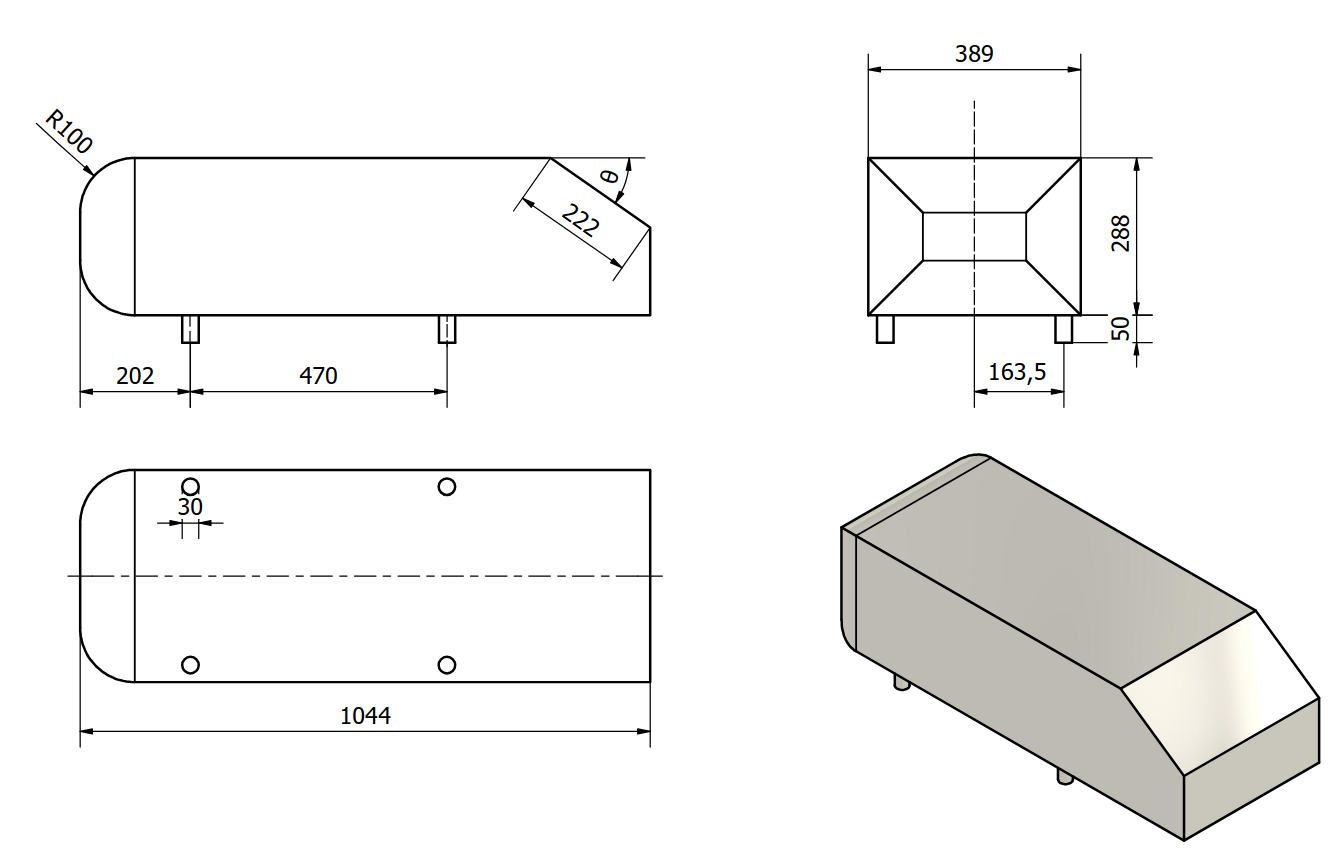}
    \captionof{figure}{Ahmed body geometry.}
    \label{fig:ahmed_geometry}
\end{center}

\begin{center}
    \includegraphics[width=0.9\linewidth]{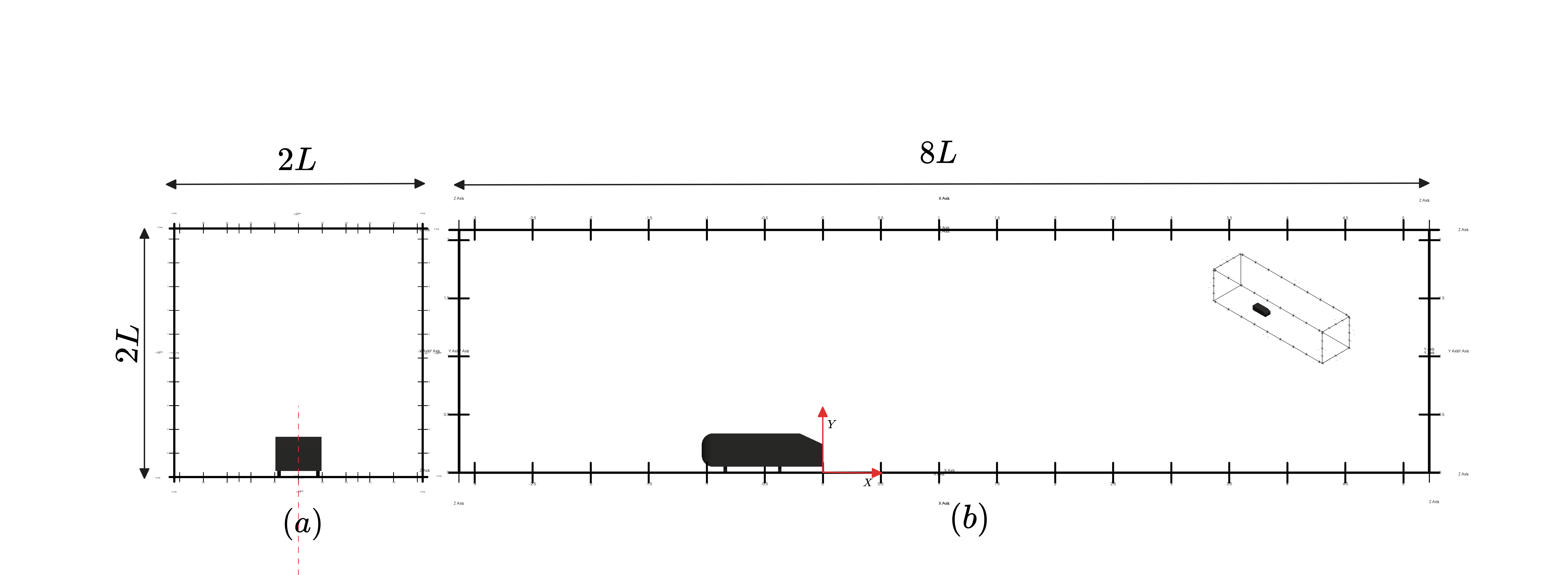}
    \captionof{figure}{Computational domain dimensions.}
    \label{fig:ahmed_domain}
\end{center}

\subsection{Mesh design, time-step sizes, and mesh-sensitivity assessment}
\label{subsec:ahmed_mesh}

The computational meshes are built with unstructured tetrahedral elements and employ local refinement near walls and in the wake, where the
dominant separated-flow physics develops(Fig.~\ref{fig:ahmed_mesh}). Refinement is concentrated around the rear slant and the downstream separated shear layers to capture boundary-layer detachment, the main recirculation bubble, and the large-scale vortical structures that govern base pressure and drag.
A smooth transition to coarser elements is applied in the far field to control the overall cost and to avoid excessive dissipation due to
abrupt mesh grading.

Three mesh levels are considered to assess the sensitivity of integral aerodynamic outputs. The coarse mesh
($\approx 3.2\times 10^6$ elements) is used to verify solver stability and obtain preliminary wake organization. The medium mesh
($\approx 1.0\times 10^7$ elements) increases resolution in the near wake and along the slant region. The fine mesh
($\approx 3.7\times 10^7$ elements) targets a resolution level appropriate for turbulence-resolving predictions of the separated wake that
governs the drag coefficient. The time-step sizes are chosen empirically to ensure stable integration and comparable unsteady behavior
across resolutions: $\Delta t = 1.0\times 10^{-4}~\mathrm{s}$, $1.0\times 10^{-5}~\mathrm{s}$, and $5.0\times 10^{-6}~\mathrm{s}$ for the
coarse, medium, and fine meshes, respectively.

Table~\ref{tab:mesh_info} reports mesh sizes together with the mean drag coefficient obtained in this work for $\theta=25^\circ$, which is the
most widely documented slant angle and therefore provides the most direct basis for quantitative comparison against reference LES and
experimental datasets \cite{SERRE201310,aljure2014turbulent}. The variation of $C_D$ between the medium and fine meshes is substantial,
highlighting the resolution demands of the high-$Re$ massively separated wake. At the same time, stable computations were obtained across all
mesh levels within a single dynamic VMS setting. Further refinement beyond the fine mesh was not pursued due to the cost of time-accurate
three-dimensional simulations with dynamic subscales. As shown in the next section, the fine-mesh value provides a competitive reference for
comparisons with turbulence-resolving results reported on meshes of comparable size. Unless otherwise stated, all subsequent Ahmed-body results
in this work refer to the fine mesh.

\begin{table}[ht!]
    \centering
    \caption{Mesh characteristics for the Ahmed-body simulations at $\theta=25^\circ$, including number of elements, mean drag coefficient obtained in this work at $\mathrm{Re} = 7.68 \times 10^{5}$, and time-step size.}
    \label{tab:mesh_info}
    \begin{tabular}{lcccc}
        \hline
        \textbf{Mesh level} & \textbf{Elements} & \textbf{Nodes} & \textbf{$C_D$ (this work, $\theta=25^\circ$)} & \textbf{$\Delta t$ [s]} \\
        \hline
        Coarse   & 3,158,028  & 590,476   & 0.4548 & $1.0 \times 10^{-4}$ \\
        Medium   & 10,132,548 & 1,979,584 & 0.4200 & $1.0 \times 10^{-5}$ \\
        Fine     & 37,283,942 & 6,082,802 & 0.3297 & $5.0 \times 10^{-6}$ \\
        \hline
    \end{tabular}
\end{table}

\begin{center}
    \includegraphics[width=0.7\linewidth]{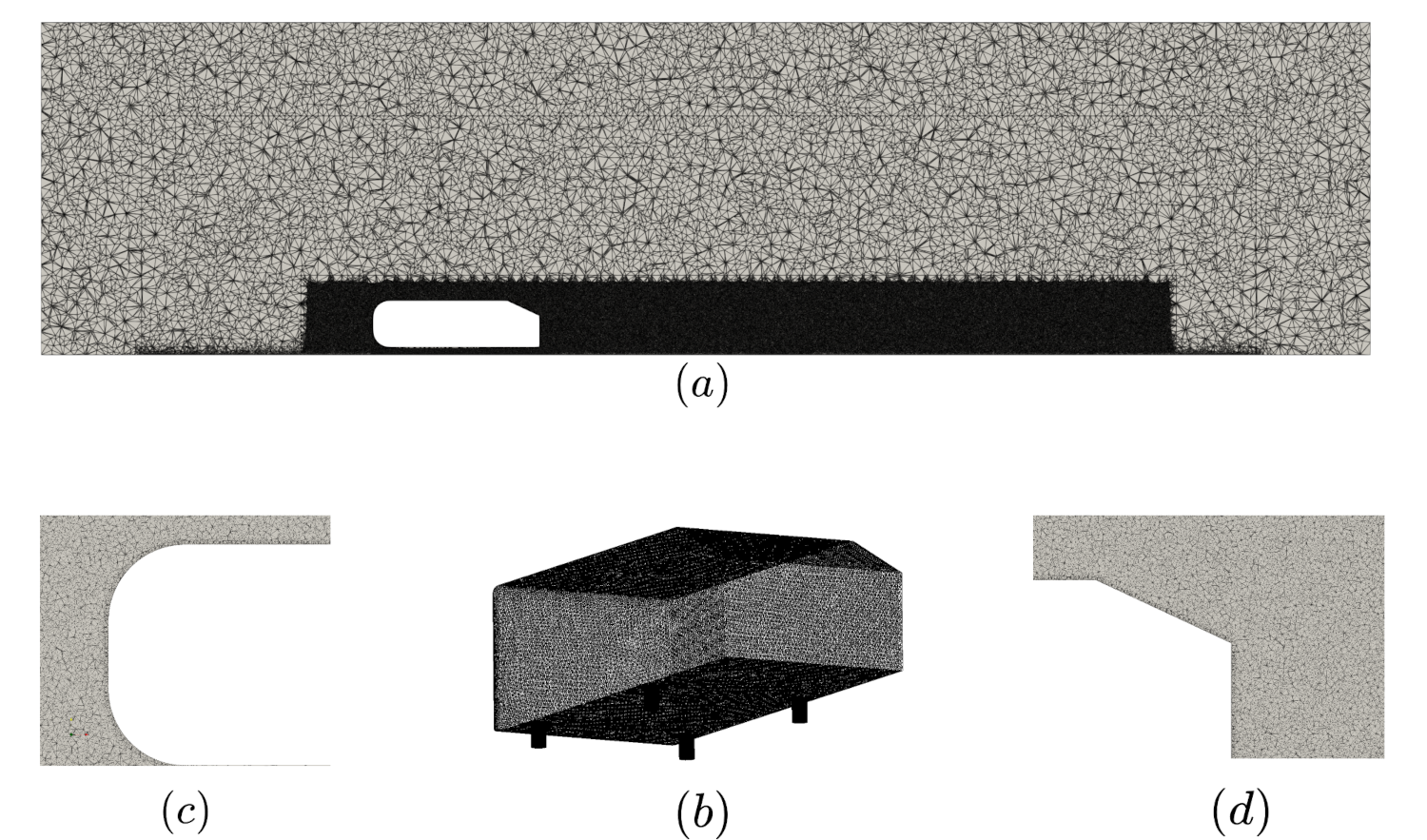}
    \captionof{figure}{Computational mesh used for the Ahmed simulations, illustrating near-wall refinement and wake resolution.}
    \label{fig:ahmed_mesh}
\end{center}

\subsection{Numerical results}
\label{subsec:ahmed_results}

Once again, the $\theta=25^\circ$ configuration is adopted as the primary quantitative validation case. The multi-angle dataset is additionally used to assess robustness across distinct wake topologies, while $\theta=25^\circ$ anchors all integral-load
comparisons.

\subsubsection{Aerodynamic coefficient and mesh sensitivity}
\label{subsubsec:ahmed_cd}

We begin with the mean drag coefficient $C_D$, which is sensitive to wake topology and near-wake pressure recovery under massive separation.
The results in Table~\ref{tab:mesh_info} show a pronounced mesh dependence at $\theta=25^\circ$: $C_D$ changes markedly between the medium mesh
(10.1M elements) and the fine mesh (37.3M elements). This sensitivity is consistent with the demanding nature of the benchmark, where accurate
integral loads require sufficient resolution of separated shear layers and of the dominant coherent structures governing base pressure.
Importantly, stable computations were obtained across all mesh levels within the proposed dynamic VMS framework, indicating robustness over a
wide range of unstructured tetrahedral resolutions.

In what follows, the fine-mesh result is taken as the reference prediction of the present method for comparisons to external data.
Table~\ref{tab:comparacion_cd_malla40} summarizes $C_D$ for $\theta=25^\circ$ against experimental measurements and representative
turbulence-resolving numerical references at comparable Reynolds numbers and mesh sizes. Our fine-mesh simulation yields
$C_D=0.3297$, within the spread of published results and close to the LES reference in Table~\ref{tab:comparacion_cd_malla40}, with a relative
difference of $4.93\%$. While experimental values exhibit variability depending on setup and corrections, the comparison indicates that the
present stabilized fractional-step formulation provides an integral-load prediction consistent with published turbulence-resolving data at
$\mathrm{Re} =\mathcal{O}(10^6)$ on practical unstructured meshes.
\begin{table}[ht!]
    \centering
    \caption{Comparison of mean drag coefficients $C_D$ for $\theta= 25^\circ$ (fine mesh used as reference for ``This work'').}
    \label{tab:comparacion_cd_malla40}
    \begin{tabular}{l|c|c|c}
        \textbf{Case} & \textbf{Elements} & \textbf{$C_D$} & \textbf{Relative diff. [\%]} \\ \hline
        This work (Fine mesh) & 37,283,942 & \textbf{0.3297} & --- \\ \hline
        Experimental & --- & 0.298 & 9.64 \\ \hline
        Guilmineau \cite{guilmineau2018hybrid} -- DES   & 23,100,000 & 0.437 & 32.55 \\
        Guilmineau \cite{guilmineau2018hybrid} -- IDDES & 23,100,000 & 0.382 & 15.83 \\
        Serre \cite{serre2013ahmed} -- LES-SVV   & 40,000,000 & 0.431 & 30.69 \\
        Serre \cite{serre2013ahmed} -- LES-NWR   & 40,000,000 & 0.346 & 4.93 \\
        Lehmkuhl \cite{lehmkuhl2019lowdissipation} -- LES & 25,600,000 & 0.297 & 9.93 \\
    \end{tabular}
\end{table}

\begin{center}
    \includegraphics[width=1\linewidth]{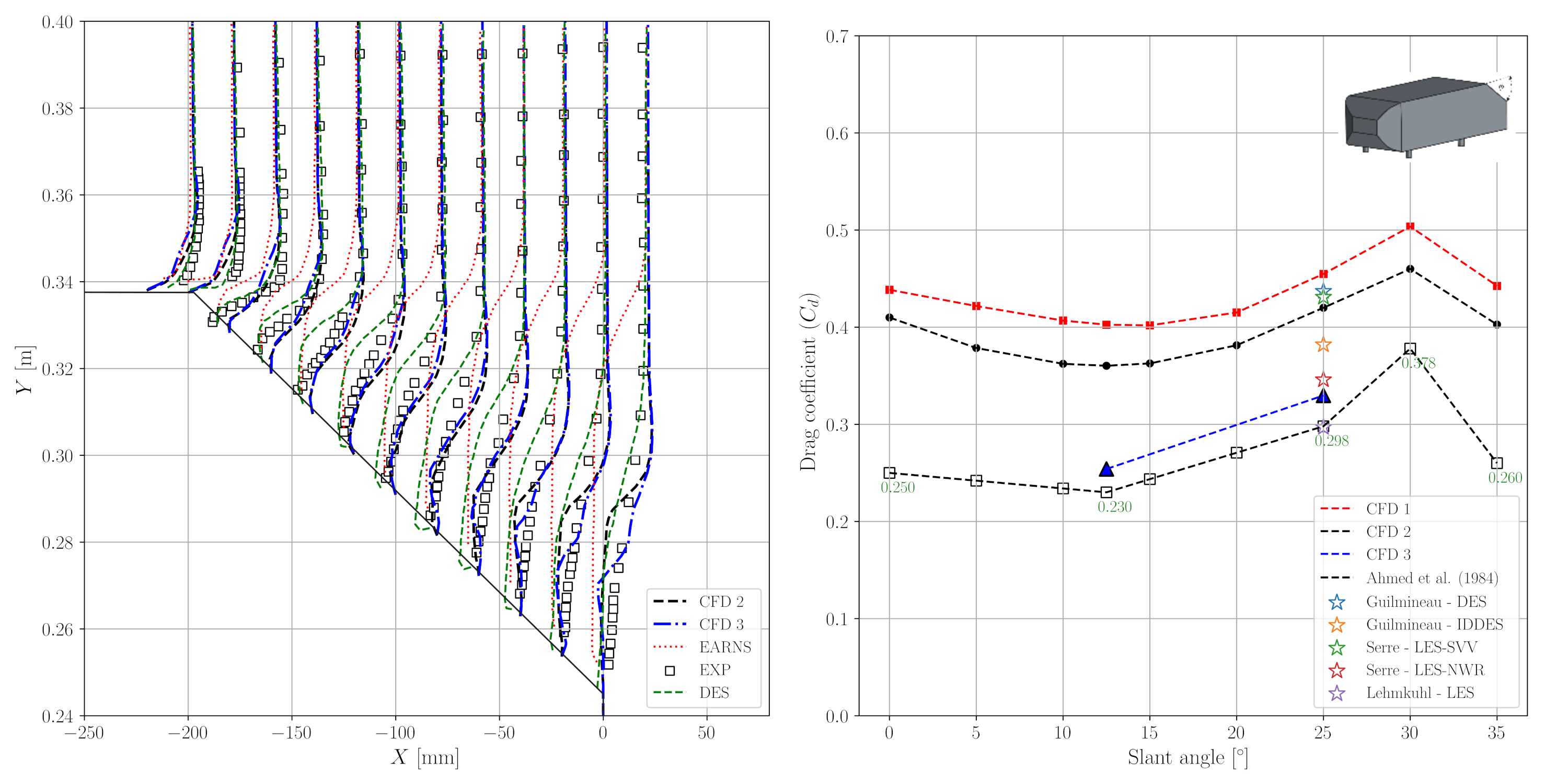}
    \captionof{figure}{Validation of aerodynamic results for the Ahmed body. (Left) Normalized mean velocity profiles at the symmetry plane for a fixed slant angle of $\theta = 25^\circ$, compared against experimental data. (Right) Variation of the mean drag coefficient $C_D$ as a function of the slant angle ($\theta$), illustrating the comparison with experimental and numerical references.}
    \label{fig:ahmed_cd_comparison}
\end{center}

Beyond the single-angle benchmark, Fig.~\ref{fig:ahmed_cd_comparison} (right) compares $C_D$ as a function of the slant angle $\theta$.
The experimental trend is well known: $C_D$ decreases from the square-back case to a minimum around $\theta\simeq 12.5^\circ$, and then
increases toward a maximum in the range $\theta\simeq 25^\circ$--$35^\circ$ due to separation on the slanted rear surface. All three mesh
levels reproduce this qualitative behavior. Quantitatively, the coarse mesh overpredicts $C_D$ across angles, while the medium mesh improves
agreement but still overestimates drag in the critical separated range. The fine mesh provides the closest agreement in this regime and is therefore adopted for all subsequent field-level analyses.

(Fig.~\ref{fig:ahmed_cd_comparison}, left) indicate that while the medium mesh captures the general wake structure, it slightly departs from the benchmarks in certain regions. In contrast, the fine mesh aligns remarkably well with the experimental measurements of \cite{ahmed1984wake}. and the numerical results by \cite{guilmineau2018hybrid}, accurately capturing both the velocity magnitude and the recovery rate at several downstream stations. This close agreement suggests that key separated-flow features are fully resolved. Overall, the observed sensitivity between 10.1M and 37.3M elements underscores both the complexity of the benchmark and the practical relevance of a stabilized formulation that remains robust across resolutions while converging toward high-fidelity, turbulence-resolving predictions.
\subsubsection{Surface pressure distribution}
\label{subsubsec:ahmed_cp}

Figure~\ref{fig:ahmed_cp_contours} shows contours of the surface pressure coefficient $C_p$ for representative slant angles. For
$\theta=0^\circ$, the pressure distribution is comparatively uniform on the rear surface, consistent with a compact near-wake region.
At $\theta=12.5^\circ$, the flow remains largely attached over the slant and the pressure field exhibits smooth gradients, correlating with
the low-drag regime. A marked change is observed at $\theta=25^\circ$, where massive separation produces an extended low-pressure region and a
strong base-pressure deficit, consistent with peak drag. At $\theta=35^\circ$, the flow remains separated but the wake reorganizes, leading to
partial pressure recovery and the characteristic decrease in $C_D$ reported beyond the critical angles.

\begin{center}
    \includegraphics[width=0.9\linewidth]{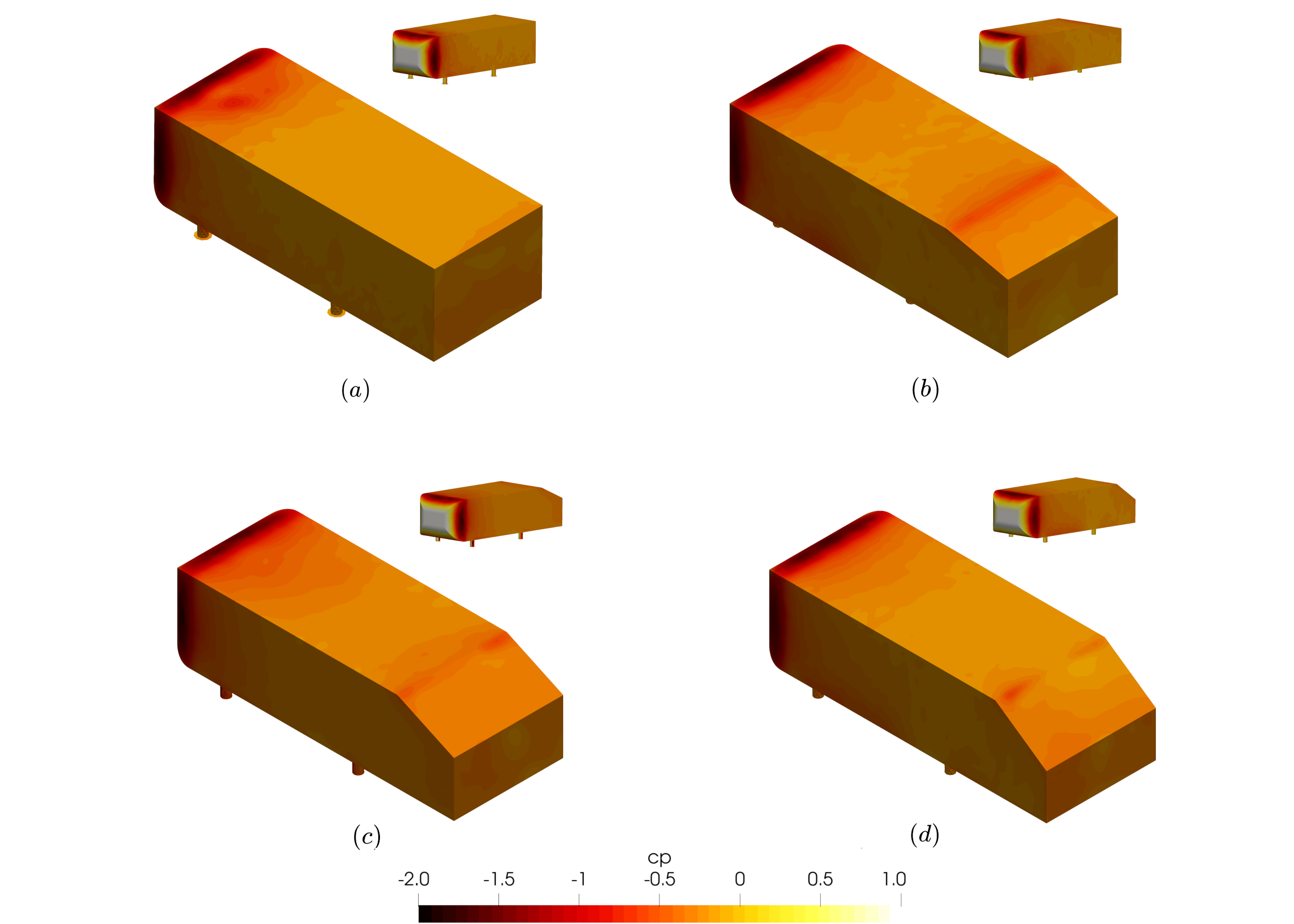}
    \captionof{figure}{Surface pressure coefficient ($C_p$) contours on the Ahmed body for different rear slant angles:
    (a) $\theta= 0^\circ$, (b) $\theta= 12.5^\circ$, (c) $\theta= 25^\circ$, and (d) $\theta= 35^\circ$.}
    \label{fig:ahmed_cp_contours}
\end{center}

\subsubsection{Wake dynamics and spectral characterization}
\label{subsubsec:ahmed_spectra}

To quantify wake unsteadiness and assess whether the simulations reproduce physically consistent inter-scale behavior, we analyze velocity and
pressure spectra from time signals sampled at monitoring locations downstream of the rear slant. The exact coordinates of these points within the control volume are specified in Table~\ref{tab:monitoring_points}; they have been positioned specifically to cover the key recirculation zones where the most significant turbulent activity occurs. Representative time histories of the three velocity components and pressure, together with  corresponding power spectra of the mean velocity and pressure and associated Lissajous diagrams used to asses phase relationships in the wake Fig.~\ref{fig:ahmed_signals}. For each monitoring signal, a discrete Fourier transform is computed and the corresponding spectral energy density is evaluated as
\begin{equation}
    E(f_k)=|\hat{u}(f_k)|^2 ,
\end{equation}
with an analogous definition for pressure. To obtain a representative wake spectrum, the individual spectra from the $M$ monitoring points are
averaged,
\begin{equation}
    \langle E(f_k)\rangle=\frac{1}{M}\sum_{m=1}^M E_m(f_k).
\end{equation}
This spatial averaging reduces pointwise intermittency and highlights dominant scaling behavior in the resolved range.

\begin{table}[h!]
    \centering
    \caption{Coordinates of monitoring points used for spectral analysis in the Ahmed-body wake.}
    \label{tab:monitoring_points}
    \begin{tabular}{cccc}
        \hline
        \textbf{Point} & \textbf{X-coordinate} & \textbf{Y-coordinate} & \textbf{Z-coordinate} \\
        \hline
        1 & 3.0   & 1.0   & 0.0      \\
        2 & 0.1   & 0.144 & 0.1945   \\
        3 & 0.3   & 0.144 & 0.389    \\
        4 & 1.0   & 0.144 & 0.0      \\
        5 & 0.5   & 0.3   & 0.0      \\
        \hline
    \end{tabular}
\end{table}

\begin{center}
    \includegraphics[width=\textwidth]{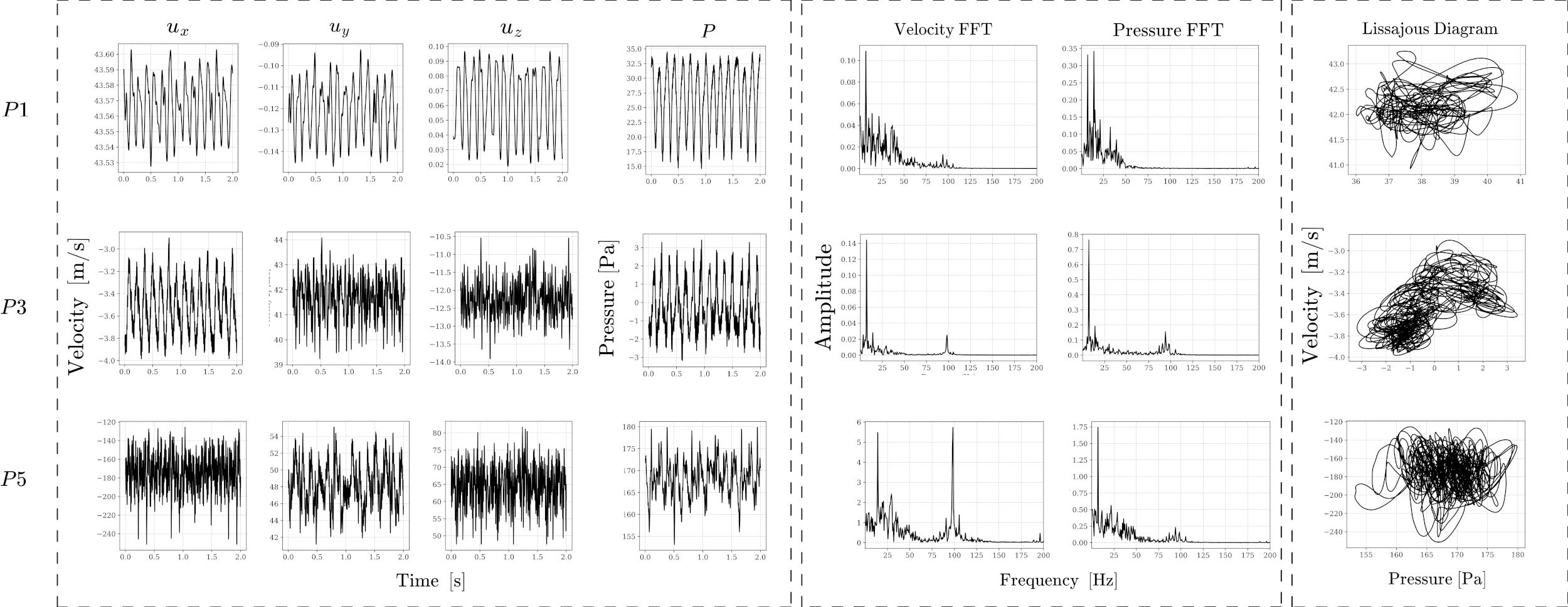}
    \captionof{figure}{Ahmed body: Representative time signals, power spectra (FFT), and phase portraits for velocity and pressure at selected monitoring points.}
    \label{fig:ahmed_signals}
\end{center}

For fully developed turbulence, inertial-range scaling is commonly associated with $E_u(f)\propto f^{-5/3}$ for velocity and, under standard
assumptions, $E_p(f)\propto f^{-7/3}$ for pressure. Figure~\ref{fig:ahmed_spectra} shows that the averaged spectra exhibit frequency ranges
compatible with these slopes over a finite interval. While the present study does not aim at a full statistical-stationarity analysis over
multiple long windows (particularly at the largest meshes), these pointwise spectral diagnostics indicate that the stabilization supplies an
appropriate dissipation behavior for turbulence-resolving simulations in massively separated external aerodynamics.

\begin{center}
    \includegraphics[width=\textwidth]{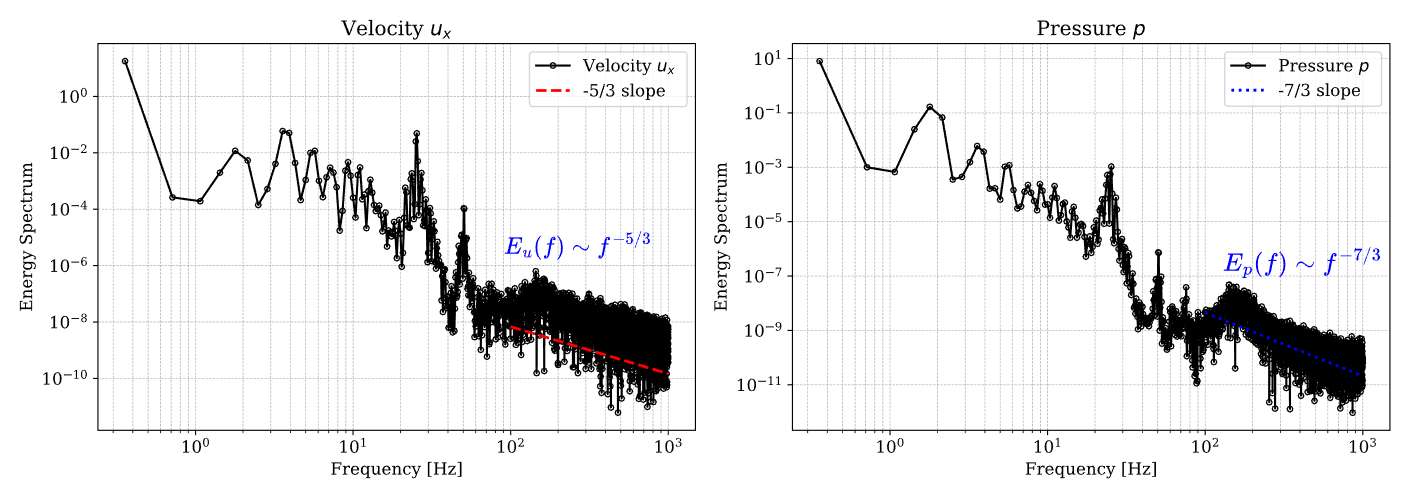}
    \captionof{figure}{Ahmed body: Averaged wake spectra for velocity (left) and pressure (right). Dashed reference lines indicate the theoretical slopes $-5/3$ and $-7/3$, respectively.}
    \label{fig:ahmed_spectra}
\end{center}

\subsubsection{Dynamic sub-scales}
Figure \ref{fig:dynamic_subscale} presents the resolved velocity and pressure fields along with the dynamic velocity subscales.The convective component $\tilde{\mathbf{u}}_1$ (bottom left) is computed from the advective residual $\rho \hat{\mathbf{u}}_h \cdot \nabla \mathbf{u}_h$, and is primarily localized within the near-wake region. On the other hand, the pressure gradient subscale $\tilde{\mathbf{u}}_2$ (bottom right) compensates for the $\nabla p_h$ residual to ensure \textit{inf-sup} stability when using equal-order elements. 

We see that the magnitude of $\tilde{\mathbf{u}}_2$ is higher than that of $\tilde{\mathbf{u}}_1$, a behavior expected in external aerodynamics where the pressure gradient is the dominant term in the momentum balance. Both subscales exhibit marked activity in the wake zone, confirming that the orthogonal projection $P_h^\perp$ restricts stabilization to the unresolved scales and preserves the energy of the macroscopic structures. The dynamic BDF2 nature of these subscales accurately captures the transient character of the flow, avoiding the excessive damping of vortex shedding frequencies typically found in quasi-static models.

\begin{center}
    \includegraphics[width=\textwidth]{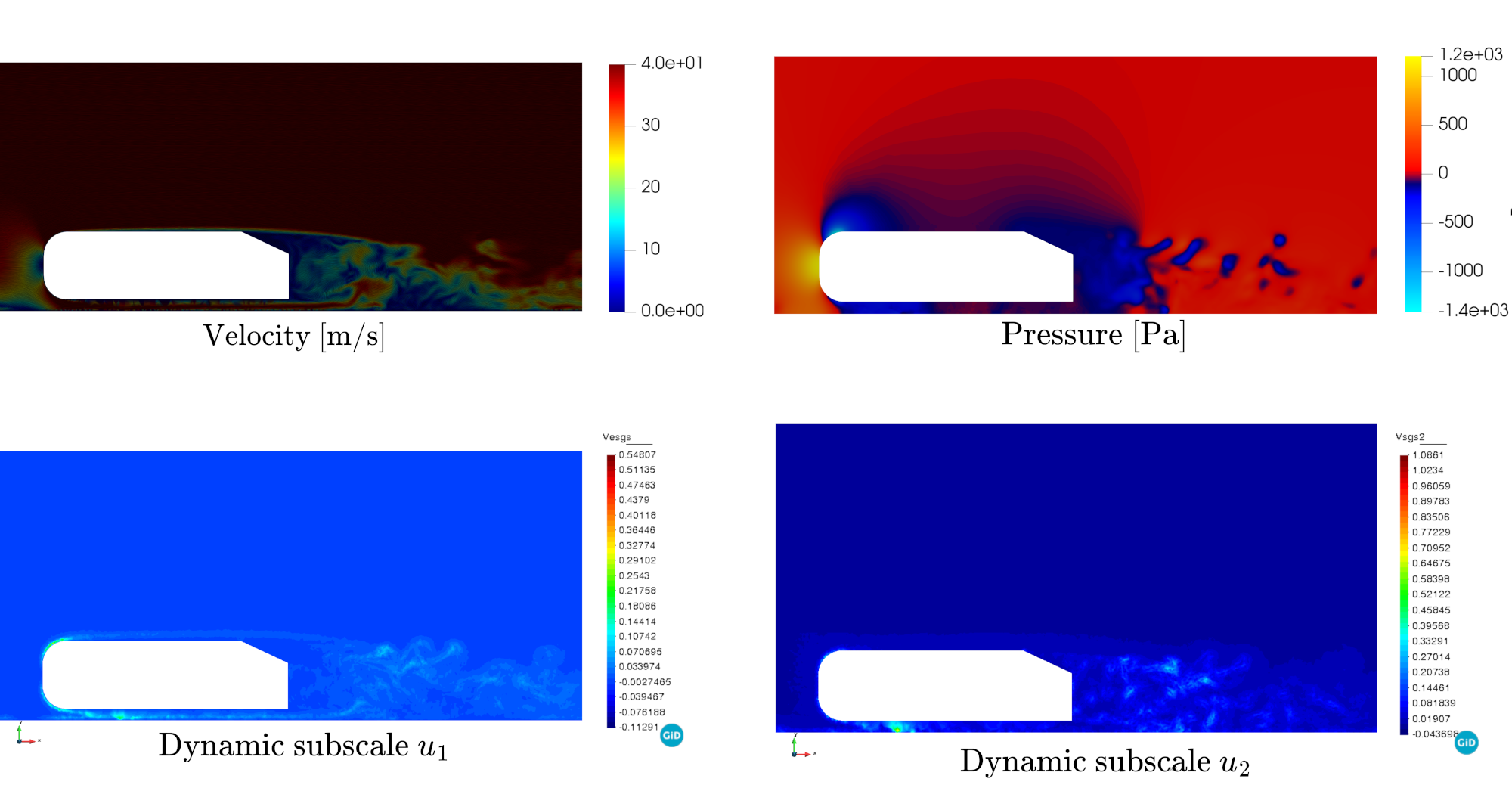}
    \captionof{figure}{Dynamic sub-scales}
    \label{fig:dynamic_subscale}
\end{center}

\subsubsection{Coherent-structure visualization and wake topology}

To visualize coherent vortical structures and clarify wake organization, we employ the $Q$-criterion,
\begin{equation}
    Q=\frac{1}{2}\left(\|\boldsymbol{\Omega}\|^2-\|\boldsymbol{S}\|^2\right),
\end{equation}
where $\boldsymbol{S}$ and $\boldsymbol{\Omega}$ are the symmetric and antisymmetric parts of the velocity gradient tensor. Positive $Q$
identifies regions where rotation dominates strain and is therefore suitable for extracting vortex cores in separated turbulent wakes.

Figure~\ref{fig:ahmed_q_criterion} shows iso-surfaces of $Q$ colored by velocity magnitude. The dominant wake topology is characterized by the
roll-up of separated shear layers and by strong streamwise vortical structures persisting over several body lengths downstream, consistent with
the base-pressure deficit associated with $C_D$. The visualization also highlights underbody-induced structures and corner vortices that interact
with the main wake and contribute to the three-dimensionality of the recirculation region. These features are known to influence wake recovery
and drag in bluff-body external aerodynamics and provide qualitative support for the resolved flow organization.

\begin{center}
    \includegraphics[width=\textwidth]{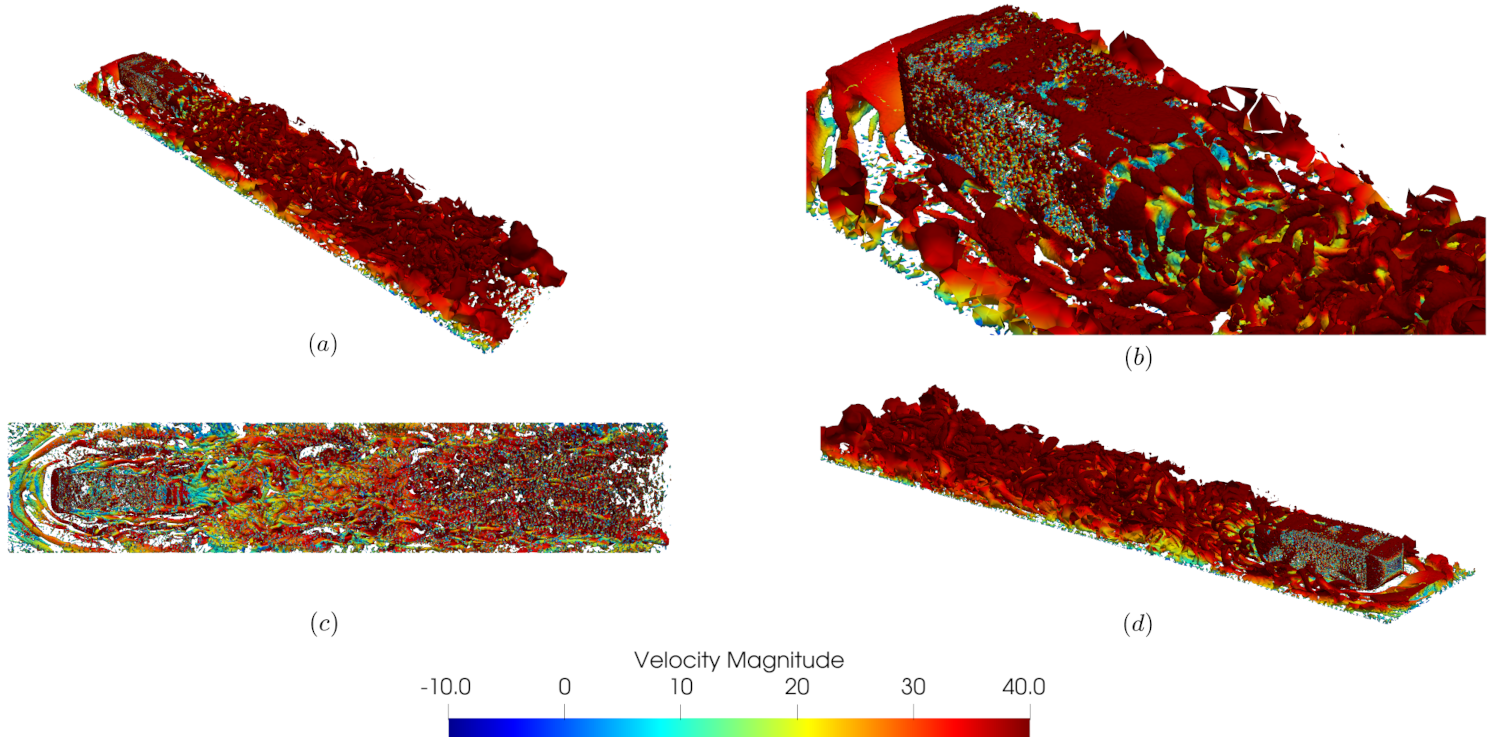}
    \captionof{figure}{Ahmed body: Iso-surfaces of the $Q$-criterion (here $Q=200$) colored by velocity magnitude.}
    \label{fig:ahmed_q_criterion}
\end{center}

Finally, Fig.~\ref{fig:ahmed_cut_planes} reports streamwise velocity contours on cross-planes downstream of the body for multiple slant angles.
The progression from attached to massively separated regimes is evident as $\theta$ increases: the recirculation region expands and the wake
deficit strengthens in the critical angles, while larger angles exhibit a more fragmented and three-dimensional wake. Complementary three dimensional streamline visualizations in Fig.~\ref{fig:ahmed_streamlines} highlight the dominant separation topology, including the formation of C-shaped vortices near the lower edges and the main vortical structure developing nature of the wake. These field-level results complement the integral-load comparison and support that the proposed stabilized fractional-step formulation delivers physically consistent turbulence-resolving predictions for the Ahmed-body benchmark.

\begin{center}
    \includegraphics[width=\textwidth]{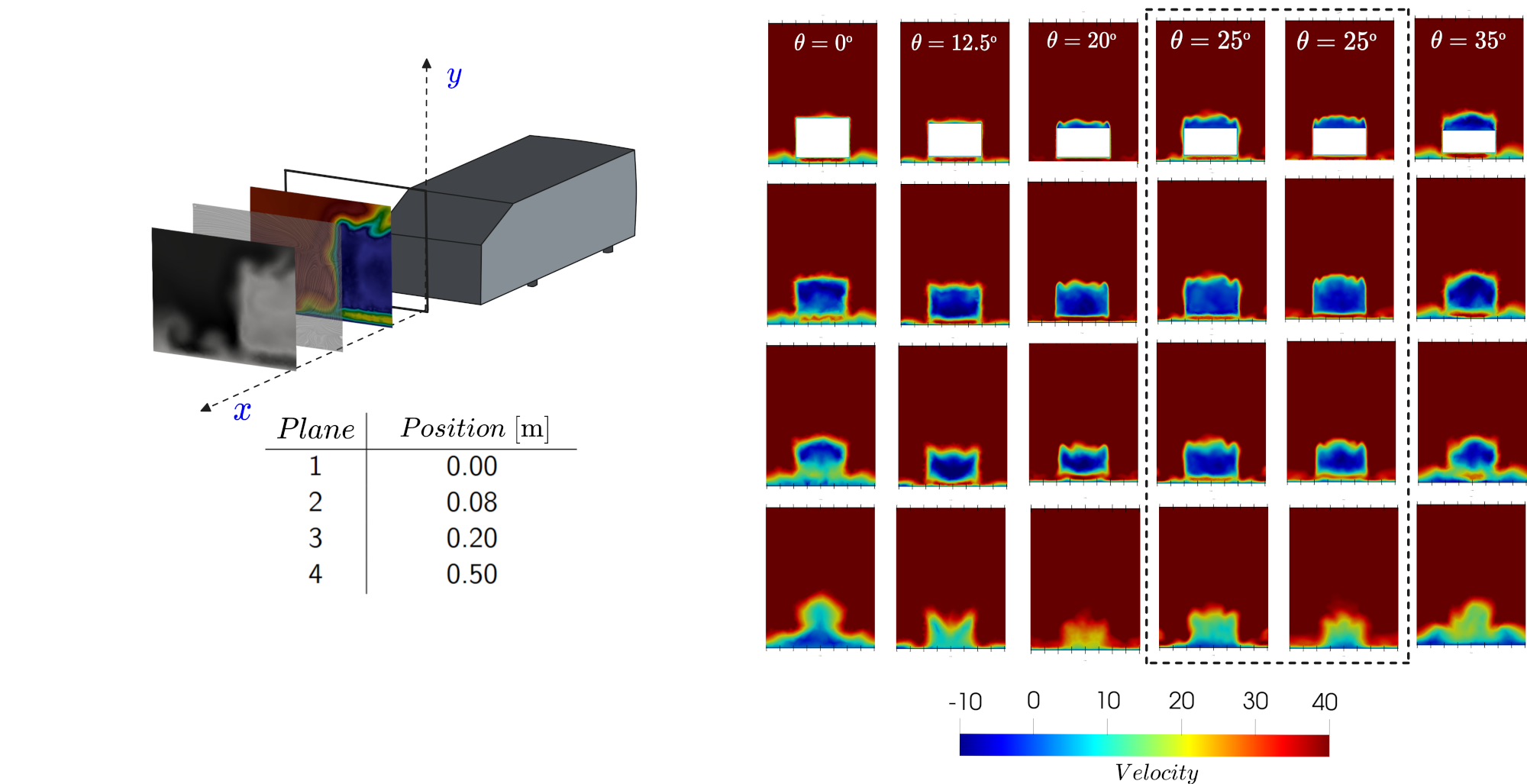}
    \captionof{figure}{Streamwise velocity contours at several downstream cross-sections for different slant angles $\theta$.}
    \label{fig:ahmed_cut_planes}
\end{center}

\begin{center}
    \includegraphics[width=\textwidth]{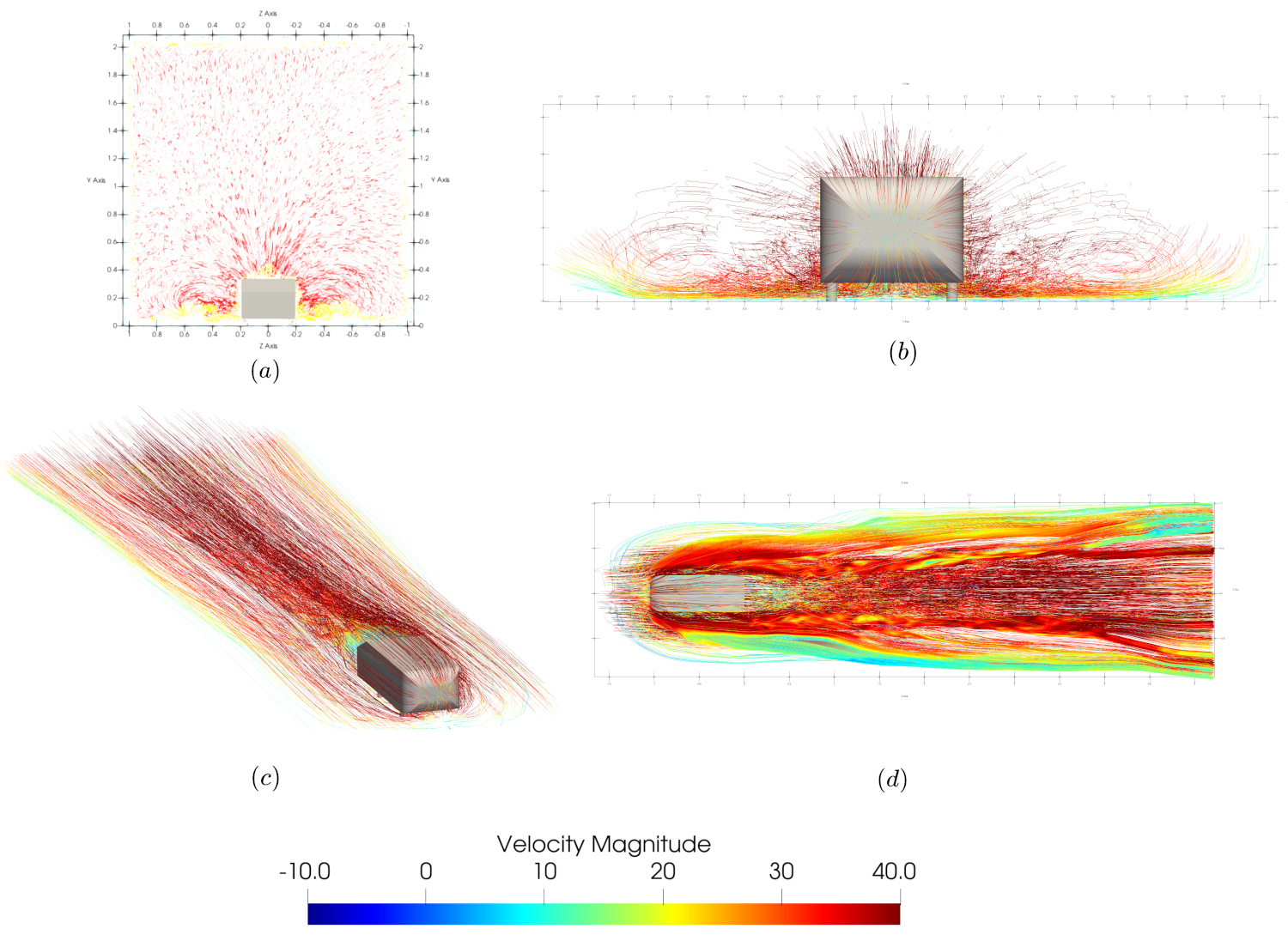}
    \captionof{figure}{Three-dimensional streamline visualization illustrating separation, shear-layer roll-up, and wake organization.}
    \label{fig:ahmed_streamlines}
\end{center}


\section{Formula~1 application}
\label{sec:f1}

To further assess the proposed stabilized fractional-step formulation in a realistic and highly demanding external-aerodynamics setting,
we consider a full-scale open-wheel Formula~1 geometry. In contrast to the Ahmed body, this configuration is not intended as a benchmark
validation case supported by a consolidated public reference database. Rather, it serves as a large-scale application that stresses the
formulation on a multi-component geometry in which aerodynamic performance depends on strongly three-dimensional interactions, ground-effect
mechanisms, and vortex-dominated flow organization.

A key aspect of this case is that the computations are performed with the proposed stabilized formulation \emph{without introducing an
explicit turbulence model}. Thus, the resolved unsteadiness and the resulting spectral content arise from geometry-induced instabilities
together with the stabilized discretization. This complements the benchmark-oriented Ahmed-body analysis by demonstrating that a purely
stabilized incompressible formulation can be advanced on a realistic configuration with complex flow topology and practical mesh sizes.

The simulation is carried out at a freestream velocity of $U_\infty = 56~\mathrm{m/s}$ ($201.6~\mathrm{km/h}$). The Reynolds number is defined
as $\mathrm{Re}=\rho U_\infty L_{\mathrm{ref}}/\mu$ using a characteristic chord length $L_{\mathrm{ref}}=0.3~\mathrm{m}$ representative of the
rear-wing elements, yielding $\mathrm{Re}\approx 1.13\cdot 10^{6}$ for the fluid properties adopted in this work. The full-scale CAD geometry used in the CDF simulations is shown in Fig .~\ref{fig:f1_geometry}. The computational domain is a wind-tunnel-like volume extending $12L$ in the streamwise direction, $3L$ laterally, and $2L$ vertically, where $L=4.9~\mathrm{m}$ is the total vehicle length (Fig.~\ref{fig:f1_domain}). The vehicle is placed over a \emph{stationary} ground plane and the tires are modeled as \emph{non-rotating} (fixed) bodies. Boundary conditions consist of a uniform inflow at the inlet, a pseudo-traction outflow at the outlet, and no-slip conditions on the vehicle and ground surfaces. Turbulent inflow fluctuations are not prescribed.

\begin{center}
    \includegraphics[width=0.75\linewidth]{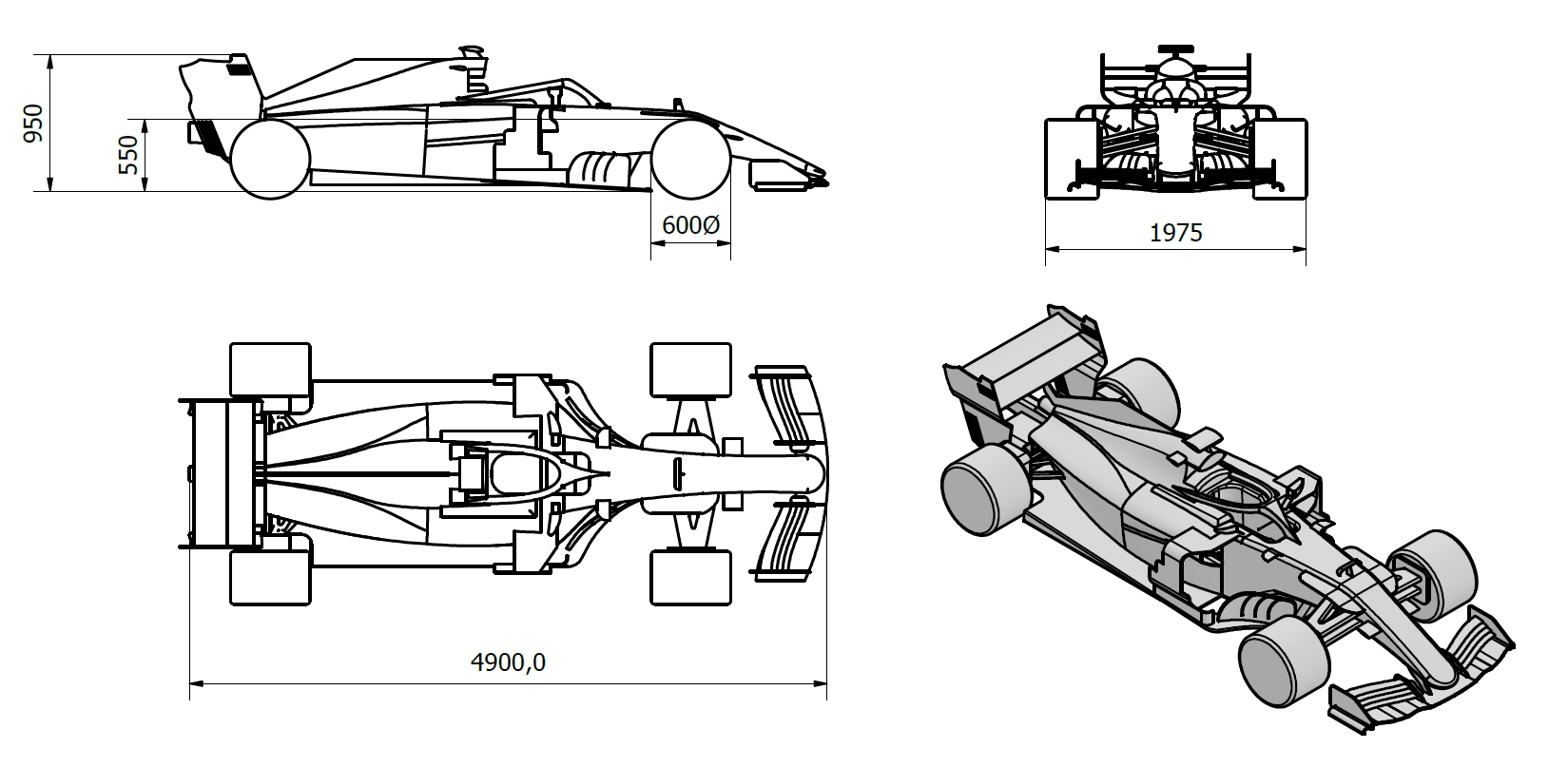}
    \captionof{figure}{CAD geometry of the full-scale Formula~1 configuration considered in this work.}
    \label{fig:f1_geometry}
\end{center}

\begin{center}
    \includegraphics[width=0.9\linewidth]{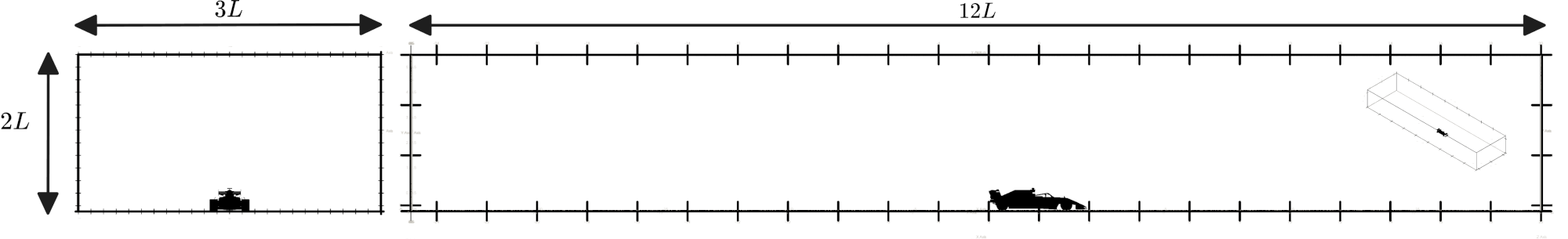}
    \captionof{figure}{Computational domain for the Formula~1 simulation (wind-tunnel-like setup).}
    \label{fig:f1_domain}
\end{center}

\paragraph{Mesh and time-step selection.}
The computational mesh is constructed with unstructured tetrahedral elements and strong local refinement near all major aerodynamic
components, including front and rear wings, underfloor, diffuser, and sidepods (Fig.~\ref{fig:f1_mesh}). Refinement targets regions where
thin shear layers and concentrated vortical structures originate and interact, while the mesh is smoothly coarsened away from the vehicle to
limit computational cost without compromising wake resolution in the near-ground region. A single fine-resolution mesh is employed,
consisting of approximately $3.87\times 10^{7}$ elements (Table~\ref{tab:mesh_f1}).

The time-step size is selected empirically to balance stability and temporal resolution of the dominant unsteady features; in this case,
$\Delta t = 2.5\times 10^{-6}~\mathrm{s}$ is adopted.

\begin{center}
    \includegraphics[width=0.7\linewidth]{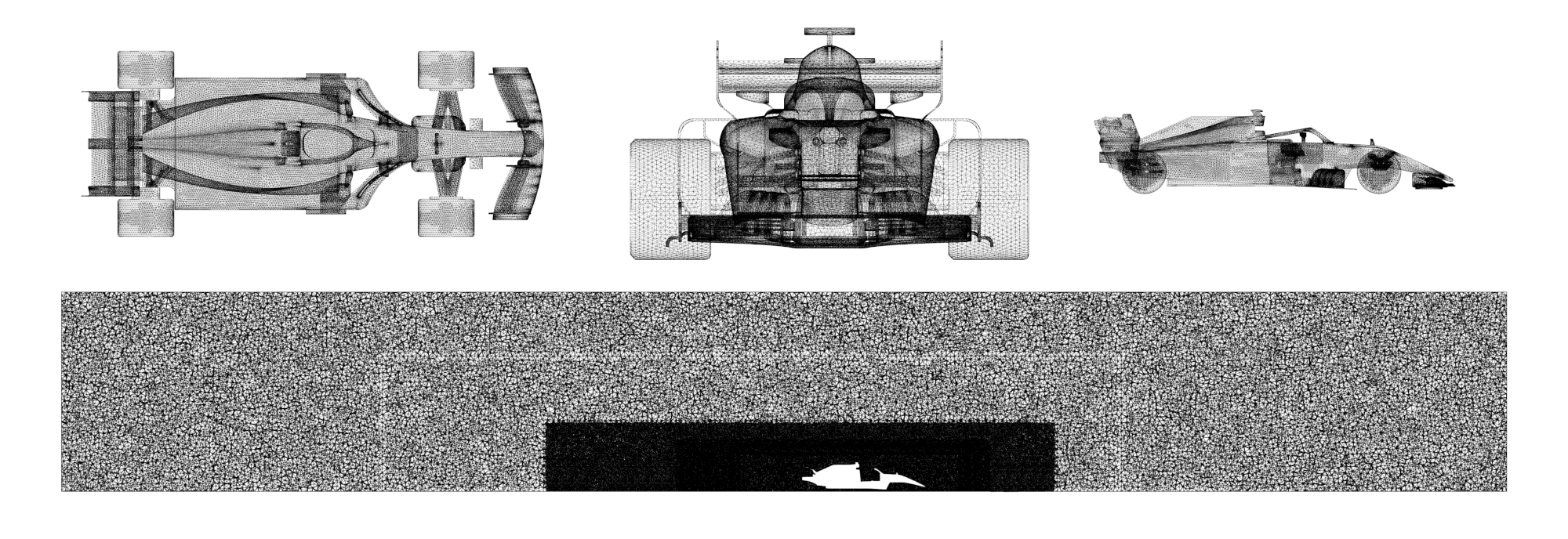}
    \captionof{figure}{Computational mesh for the Formula~1 simulation, highlighting near-body refinement and wake resolution.}
    \label{fig:f1_mesh}
\end{center}

\subsection{Numerical results}
\label{subsec:f1_results}

\begin{table}[ht!]
    \centering
    \caption{Mesh characteristics and integral aerodynamic coefficients for the Formula~1 simulation.}
    \label{tab:mesh_f1}
    \begin{tabular}{lccccccc}
        \hline
        \textbf{Mesh} & \textbf{Elements} & \textbf{$L$ [m]} & \textbf{$Re$} & \textbf{$C_D$} & \textbf{$C_L$} & \textbf{$C_L/C_D$} & \textbf{$\Delta t$ [s]} \\
        \hline
        Fine & 38,728,716 & 4.9 & $1.13 \times 10^6$ & 1.56 & -4.69 & -3.00 & $2.5 \times 10^{-6}$ \\
        \hline
    \end{tabular}
\end{table}

\paragraph{Integral aerodynamic coefficients.}
Table~\ref{tab:mesh_f1} reports the drag coefficient $C_D$, the lift coefficient $C_L$ (negative values correspond to downforce), and the
ratio $C_L/C_D$. These coefficients are computed from the total aerodynamic force evaluated on the vehicle surface at $U_\infty=56~\mathrm{m/s}$
using a reference area $A=1.5~\mathrm{m}^2$. The corresponding mean forces are $F_D = 4\, 513~\mathrm{N}$ and $F_L = -13\,586~\mathrm{N}$.
Given the application-oriented nature of this case, emphasis is placed on the resolved flow organization and on internal consistency
diagnostics (coherent-structure topology and spectra), rather than on comparison to a specific public reference dataset.

\paragraph{Cross-sectional wake evolution.}
Figure~\ref{fig:f1_cutplanes} visualizes the streamwise development of the flow using transverse planes colored by velocity magnitude. The
near field exhibits multiple interacting low-velocity regions associated with wakes from exposed wheels and upstream lifting surfaces. As the
flow progresses downstream, these wake structures merge and reorganize into a broadened and strongly three-dimensional wake. Further
downstream, the wake displays a pronounced vertical displacement and cross-sectional deformation, consistent with the combined action of
underbody acceleration, diffuser-induced expansion, and rear-wing upwash in a ground-effect configuration.

\begin{center}
    \includegraphics[width=\textwidth]{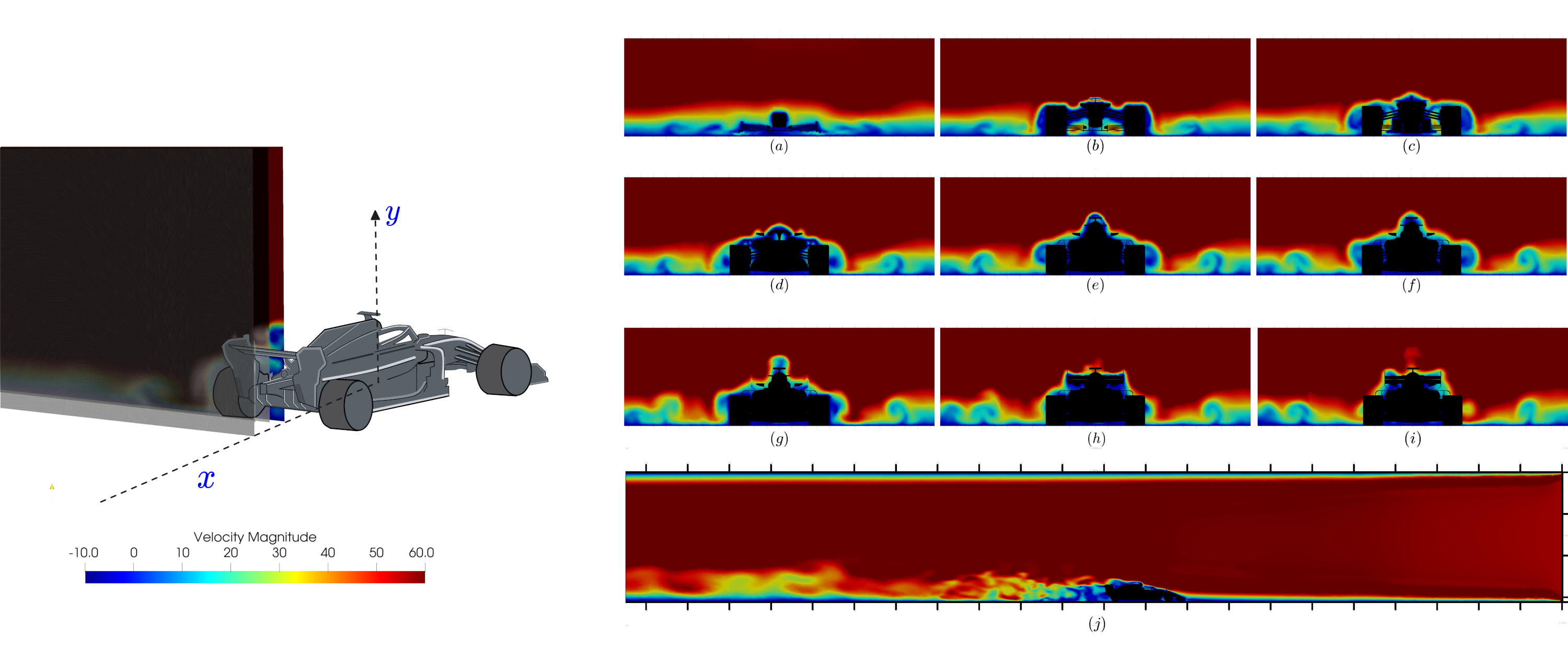}
    \captionof{figure}{Streamwise evolution of the wake visualized on transverse cut planes colored by velocity magnitude.}
    \label{fig:f1_cutplanes}
\end{center}

\paragraph{Surface pressure distribution.}
The surface pressure coefficient contours in Fig.~\ref{fig:f1_pressure} identify dominant load-bearing regions. High-pressure zones are
observed at stagnation regions on the nose and on the frontal faces of the tyres, whereas extended low-pressure regions develop on the suction
sides of the front and rear wings and along the underbody. The resulting pressure field is consistent with downforce generation through
large-scale pressure differences, and provides mechanically relevant loading signatures for potential coupled analyses.

\begin{center}
    \includegraphics[width=0.75\textwidth]{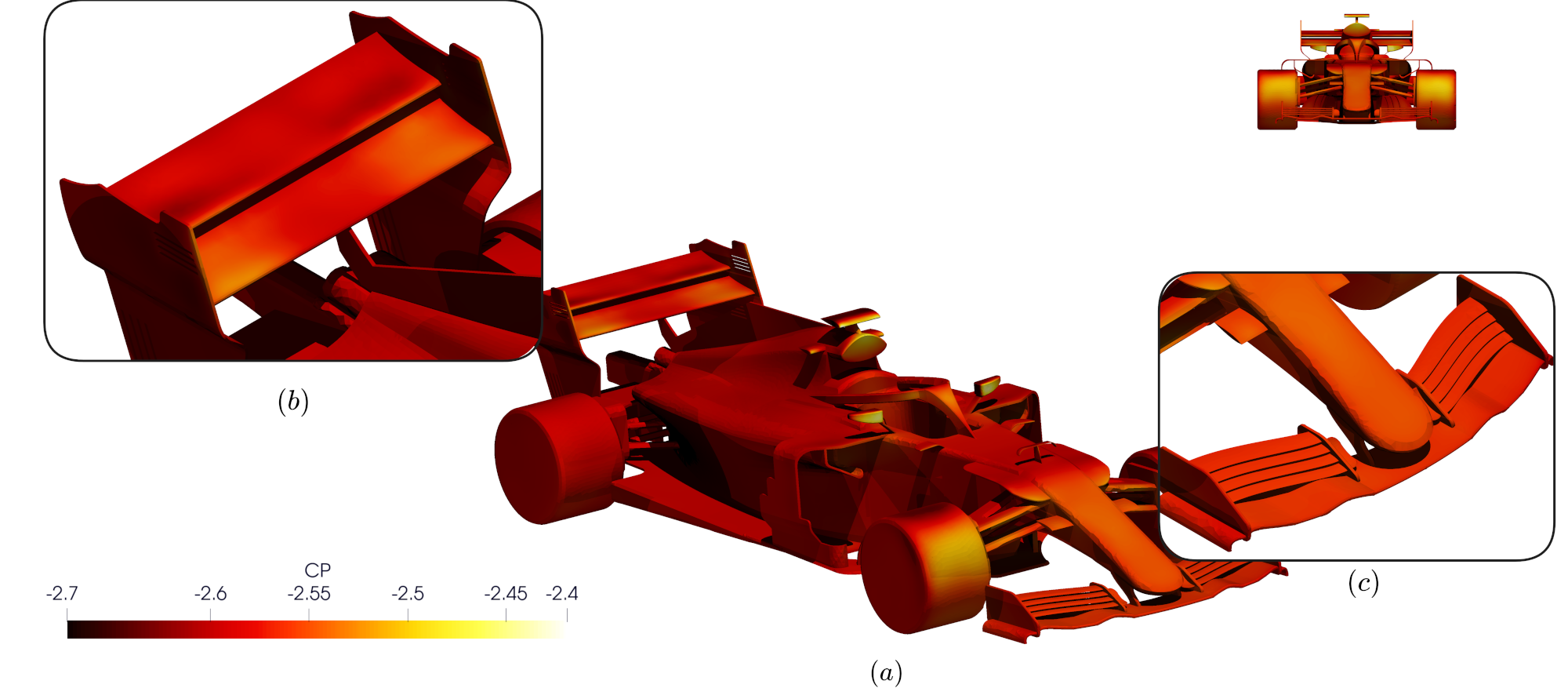}
    \captionof{figure}{Surface pressure coefficient ($C_p$) contours for the Formula~1 configuration.}
    \label{fig:f1_pressure}
\end{center}

\paragraph{Coherent vortical structures (Q-criterion).}
Figure~\ref{fig:f1_q_criterion} reports iso-surfaces of the $Q$-criterion colored by velocity magnitude. The visualization reveals interacting
vortical structures generated by wheel wakes and by multi-element lifting devices. Elongated vortex tubes are observed downstream of the
wheels and along the vehicle flanks, and a strong vortical organization is also visible in the underbody and diffuser region. The resulting
topology highlights the strongly three-dimensional and multi-scale nature of the flow in this configuration.

\begin{center}
    \includegraphics[width=0.75\textwidth]{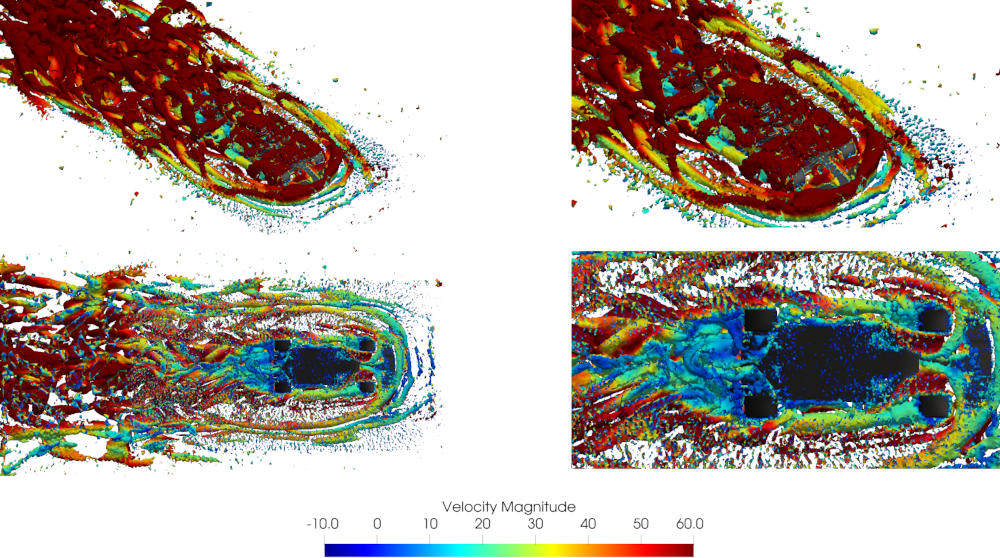}
    \captionof{figure}{F1 car: Iso-surfaces of the $Q$-criterion colored by velocity magnitude.}
    \label{fig:f1_q_criterion}
\end{center}

\paragraph{Streamline visualization.}
The streamline views in Fig.~\ref{fig:f1_streamlines} complement the $Q$-criterion by showing global transport pathways. Near the front,
streamlines deviate laterally around the wheel region and reorganize around the front aerodynamic devices. Along the sidepods and upper
bodywork, acceleration and curvature effects lead to streamline compression and subsequent expansion downstream. At the rear, streamlines show
interaction between the underbody region and the rear lifting system, contributing to wake deflection and recovery patterns consistent with
ground-effect aerodynamics.

\begin{center}
    \includegraphics[width=\textwidth]{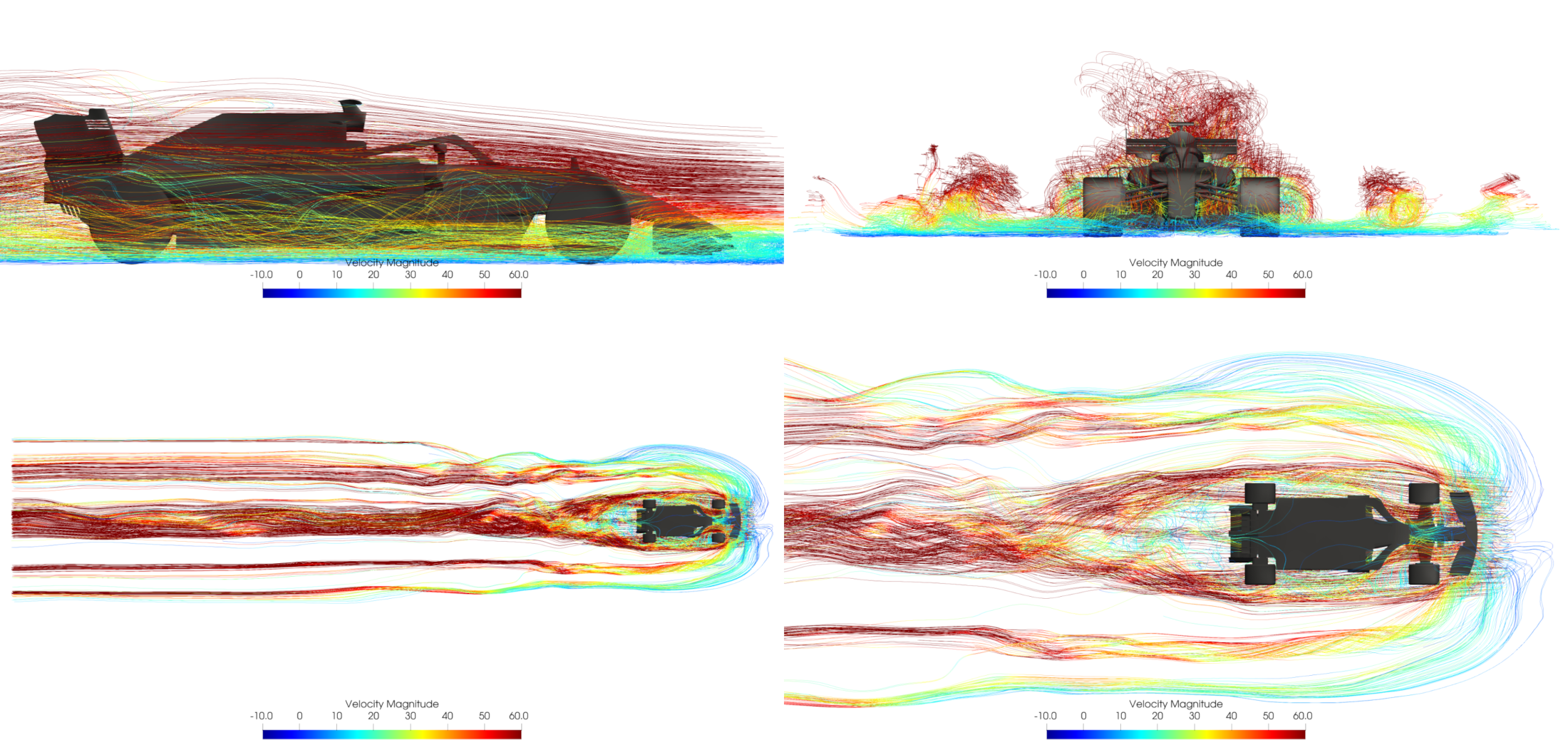}
    \captionof{figure}{Three-dimensional streamline visualization for the Formula~1 flow.}
    \label{fig:f1_streamlines}
\end{center}

\paragraph{Wake unsteadiness and spectral signatures.}
To assess inter-scale behavior in the resolved dynamics, monitoring points are placed in the near and far wake. Their exact coordinates within the control volume are detailed in Table~\ref{tab:monitoring_points_f1}; these locations are strategically chosen to capture the dynamics within the primary recirculation zones.

\begin{table}[h!]
    \centering
    \caption{Coordinates of monitoring points used for spectral analysis in the F1 model.}
    \label{tab:monitoring_points_f1}
    \begin{tabular}{cccc}
        \hline
        \textbf{Point} & \textbf{X [m]} & \textbf{Y [m]} & \textbf{Z [m]} \\
        \hline
        1 & -2.5 & 0.25 & 0.1  \\
        2 & -0.6 & 0.05 & 0.2  \\
        3 & -4.0 & 0.3  & 0.15 \\
        4 & -2.5 & 0.25 & -0.1 \\
        5 & -0.6 & 0.05 & -0.2 \\
        \hline
    \end{tabular}
\end{table}

Figure~\ref{fig:f1_signals}
reports representative velocity fluctuations in time. In particular, the lower mean velocity magnitudes observed in certain signals correspond to locations within the recirculation bubbles, where large-scale vortices are generated and shed. The corresponding energy spectral densities are shown in Fig.~\ref{fig:f1_spectra} for
both velocity and pressure. The spectra exhibit frequency ranges compatible with classical inertial-subrange slopes, namely
$E_u(f)\propto f^{-5/3}$ for velocity and $E_p(f)\propto f^{-7/3}$ for pressure. These spectra provide an a posteriori consistency diagnostic
for the resolved wake dynamics in this under-resolved regime.

\begin{center}
    \includegraphics[width=\textwidth]{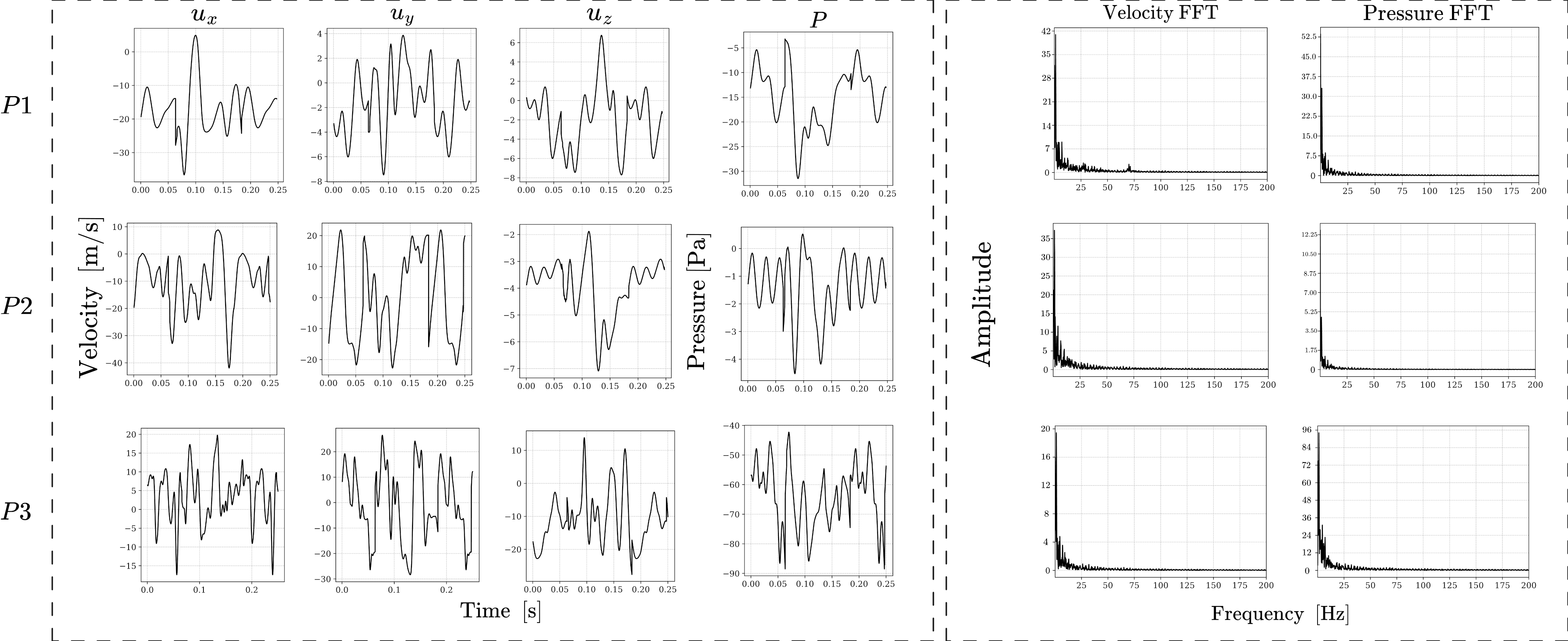}
    \captionof{figure}{F1 car: Representative velocity fluctuations at wake monitoring points.}
    \label{fig:f1_signals}
\end{center}

\begin{center}
    \includegraphics[width=\textwidth]{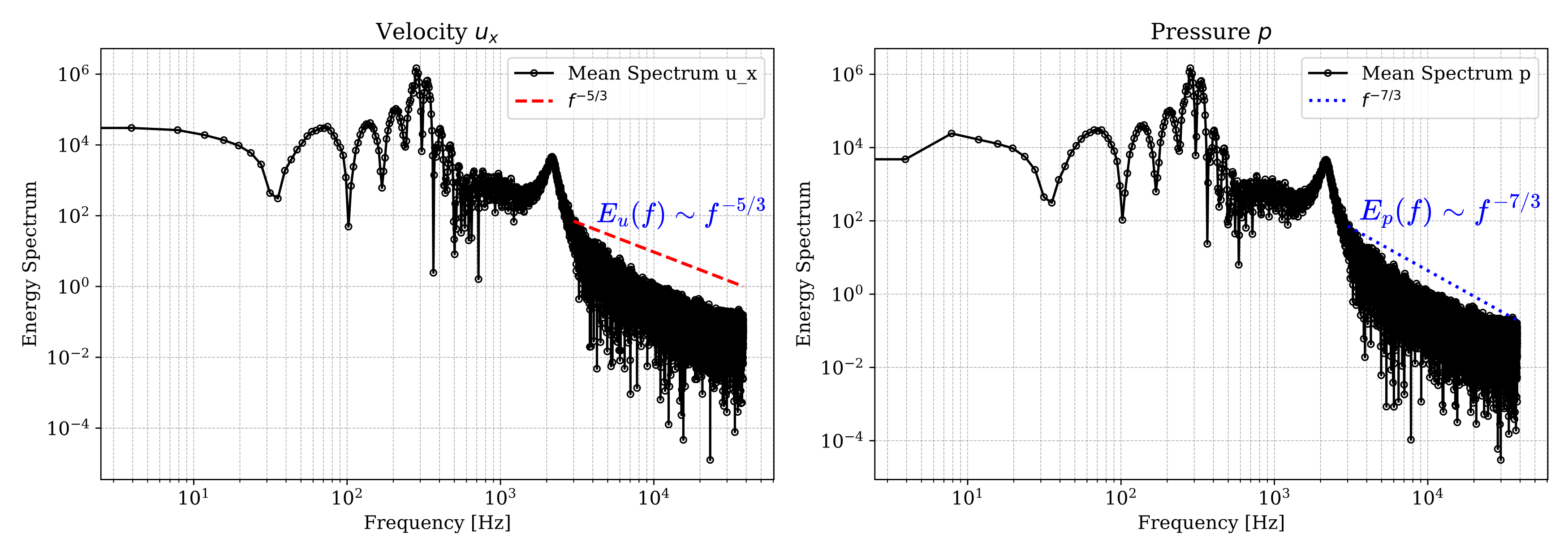}
    \captionof{figure}{F1 car: Energy spectra for velocity (left) and pressure (right) at wake monitoring points. Dashed lines indicate the $-5/3$ and $-7/3$ slopes.}
    \label{fig:f1_spectra}
\end{center}

\section{Conclusion}
\label{Conclusion}

This work proposed and assessed a dynamic, term-by-term variational multiscale (VMS) stabilization tailored to an incremental
pressure-correction fractional-step framework for incompressible flows ranging from laminar to turbulent regimes.
The formulation is designed at the fully discrete monolithic level and then embedded into the projection stages by modifying existing
operator blocks (rather than introducing additional velocity--pressure cross couplings). This preserves the computational organization
of the fractional-step method while introducing additional dissipation through orthogonal dynamic subscales. The main conclusions drawn from our study are:
\begin{itemize}
  \item A minimal, term-by-term VMS-inspired stabilization with \emph{orthogonal} projections and \emph{dynamic} subscales can be integrated
  into an incremental pressure-correction scheme without altering its stage-wise structure. In block form, the stabilization contributions
  act as (i) momentum-side modifications entering the velocity predictor and (ii) a pressure-side diagonal augmentation entering the
  pressure-correction (Poisson-like) stage, while subscale-history terms appear only as additional known right-hand-side contributions.
  This localization retains the efficiency and modularity of projection methods in large-scale computations.

  \item The method supports equal-order velocity--pressure interpolation on large unstructured tetrahedral meshes and remains robust in
  convection-dominated regimes relevant to external aerodynamics. The orthogonal term-by-term construction yields a clear stabilized layout
  in which each contribution can be associated with a specific operator block, enabling targeted assessment of robustness and dissipation
  effects.

  \item Validation on the Ahmed-body benchmark at $\mathrm{Re} =7.68\times 10^{5}$ showed that the proposed stabilized fractional-step formulation
  captures massively separated wake dynamics and delivers competitive integral-load predictions across multiple rear slant angles.
  The $\theta=25^\circ$ configuration, adopted as the primary quantitative reference due to the breadth of available literature, exhibited
  pronounced mesh sensitivity between the medium ($\approx 10$M elements) and fine ($\approx 37$M elements) meshes, underscoring the
  resolution demands of this benchmark. Importantly, the formulation remained numerically stable across all mesh levels considered and
  converged toward a fine-mesh drag coefficient consistent with reference turbulence-resolving results at comparable mesh sizes.

  \item Spectral diagnostics based on wake monitoring signals provided an a posteriori consistency indicator for dissipation behavior: the
  computed velocity and pressure spectra exhibited frequency ranges compatible with inertial-subrange scaling. These spectra complement
  integral coefficients and field visualizations as practical indicators in large-scale separated external-aerodynamics computations.

  \item Applicability was further demonstrated on a realistic full-scale Formula~1 configuration at $U_\infty=56~\mathrm{m/s}$ (201.6~km/h) and
  $\mathrm{Re}\approx 1.13\times 10^{6}$, using a single fine unstructured mesh of approximately $4\times 10^{7}$ tetrahedral elements. This case is
  not used as a benchmark validation against a consolidated public database; instead, it stresses robustness on a multi-component geometry
  with strong three-dimensional interactions, ground-effect mechanisms, and vortex-dominated flow organization. The computations were advanced
  without introducing an explicit turbulence model, so robustness and dissipation relied on the proposed stabilized formulation alone,
  illustrating its versatility for practically relevant configurations.

  \item From an algorithmic standpoint, the optional grad--div contribution acting through the pressure subscale improved
  nonlinear robustness in the most demanding configuration (Formula~1). With this term switched on, the nonlinear iterations converged on the available meshes and time-step sizes; without it, the iterations tended to stagnate, suggesting that additional stabilization can be beneficial in the most demanding settings.

  \item The choice of stabilization parameters was practically relevant at scale. In particular, adopting $c_3=24$ in the definition of
  $\tau_2$ (departing from common choices where $c_3=c_1$) provided the most robust nonlinear behavior across the reported test cases and
  improved convergence in the Formula~1 computation. This choice was guided by numerical experimentation after establishing a convergent
  baseline for the Formula~1 case and was retained throughout the study.
\end{itemize}

Future work may consider extensions that improve efficiency and spectral fidelity in large external-aerodynamics settings, including
mesh adaptivity ($h$- or $r$-adaptivity) to concentrate resolution in separation regions and dominant vortex cores, higher-order finite
elements, alternative time integrators (e.g., higher-order multistep schemes such as BDF3) to enable larger time steps at comparable
accuracy, and enhanced nonlinear solvers (Newton or hybrid Picard--Newton strategies and fixed-point acceleration such as Anderson
acceleration). As secondary improvements, longer sampling windows and additional statistics (component-wise spectra, correlation functions,
and integrated energy budgets) could strengthen turbulence characterization, and multiphysics extensions such as aeroelasticity/FSI suggested
by the computed loads are natural directions best addressed in dedicated follow-on studies.

\section*{Acknowledgements}
D. Escobar acknowledges the financial support of ANID-Chile, through the Scholarship program / MAGÍSTER-NACIONAL / 2023-22230392. A. Aguirre acknowledges the financial support of Dirección de Investigación Científica y Tecnológica (Dicyt) of the University of Santiago of Chile through the project DICYT-052516AR-REG. D. Pacheco acknowledges funding by the Federal Ministry of Education and Research (BMBF) and the Ministry of Culture and Science of the German State of North Rhine-Westphalia (MKW) under the Excellence Strategy of the Federal Government and the Länder. E. Castillo  appreciate the financial support granted by the National Research and Development Agency (ANID) of Chile, through the project FONDECYT 1250287.

\bibliographystyle{unsrtnat}

\addcontentsline{toc}{section}{references}
\bibliography{References}
\end{document}